\newcommand{\reviewformat}{%
  \documentclass[sigconf]{acmart}
}

\newcommand{\conferenceformat}{%
  \documentclass[sigconf]{acmart}
}

\reviewformat
\AtBeginDocument{%
  
    \setlength{\parskip}{0pt}
    }

\setcopyright{acmlicensed}
\copyrightyear{2026}
\acmYear{2026}
\acmDOI{XXXXXXX.XXXXXXX}
\acmConference[ACM UIST '26]{UIST 2026}{November 2--November 5, 2026}{Detroit, Michigan}
\acmISBN{978-1-4503-XXXX-X/2018/06}

\newcommand{\hl}[2]{{\color{#1}\bfseries #2}}
\usepackage{tabularx} 
\usepackage{booktabs} 
\usepackage[ruled]{algorithm2e} 
\usepackage[colorinlistoftodos]{todonotes}
\usepackage{verbatim}
\usepackage{graphicx}
\usepackage{xspace}
\usepackage{tikz}
\usepackage{wrapfig}
\usepackage{colortbl}  
\usepackage{xcolor}    
\usepackage{tcolorbox}

\newtheoremstyle{tightdef}%
  {0.3em}   
  {0.3em}   
  {\normalfont} 
  {}        
  {\itshape} 
  {.}       
  {0.5em}   
  {}        

\theoremstyle{tightdef}
\newtheorem{definition}{Definition}[section]

\newcommand{\defterm}[1]{\textit{#1}}

\graphicspath{ {./figures/} }
\usetikzlibrary{shapes.geometric, arrows, positioning}

\tikzset{
    startstop/.style = {rectangle, rounded corners, minimum width=3cm, minimum height=1cm, text centered, draw=black, fill=red!30},
    process/.style = {rectangle, minimum width=3cm, minimum height=1cm, text centered, draw=black, fill=blue!20},
    arrow/.style = {thick,->,>=stealth}
}

\usepackage{balance}       
\usepackage[T1]{fontenc}   
\usepackage{hyperref}
\usepackage{color}
\usepackage{booktabs}

\usepackage{microtype}        
\usepackage{ccicons}          

\usepackage{cuted}
\usepackage{capt-of}

\usepackage{upquote}
\usepackage{listings}
\usepackage{float}
\newfloat{infobox}{h}{lob}
\floatname{infobox}{Box}

\usepackage{amsmath}

\definecolor{codegreen}{rgb}{0,0.6,0}
\definecolor{codegray}{rgb}{0.5,0.5,0.5}
\definecolor{codepurple}{rgb}{0.58,0,0.82}

\lstdefinestyle{mystyle}{
    backgroundcolor=\color{white},
    commentstyle=\color{codegreen},
    keywordstyle=\color{magenta},
    numberstyle=\tiny\color{codegray},
    stringstyle=\color{codepurple},
    basicstyle=\ttfamily\footnotesize,
    breakatwhitespace=false,
    breaklines=true,
    captionpos=b,
    keepspaces=true,
    numbers=left,
    numbersep=5pt,
    showspaces=false,
    showstringspaces=false,
    showtabs=false,
    tabsize=2
}

\definecolor{light-gray}{gray}{0.95}

\definecolor{lightgray}{rgb}{0.83, 0.83, 0.83}
\definecolor{melon}{rgb}{0.99, 0.74, 0.71}

\usepackage{multirow}

\usepackage{blindtext}
\usepackage{subcaption}
\usepackage{caption}
\usepackage{enumitem}
\usepackage{graphicx}
\usepackage{multicol}

\usepackage{appendix}

\usepackage{tikz}
\usetikzlibrary{positioning,arrows,graphs,backgrounds,fit,calc,quotes,shapes.multipart}
\usepackage{pgfplots}
\usepackage{pgfplotstable}

\pgfplotsset{width=3.5cm,
    tick label style={font=\footnotesize}
}

\definecolor{tisanecodetop}{RGB}{74,63,106}
\definecolor{closecolor}{RGB}{225,93,87}
\definecolor{minimizecolor}{RGB}{245,215,90}
\definecolor{maximizecolor}{RGB}{111,208,74}
\usepackage{wrapfig}

\SetAlFnt{\small}
\SetAlCapFnt{\small}
\SetAlCapNameFnt{\small}
\SetAlCapHSkip{0pt}
\IncMargin{-\parindent}

\newcommand{\eunice}[1]{\hl{orange}{EJ: #1}}
\newcommand{\london}[1]{\hl{olive}{LB: #1}}

\newcommand{\keyidea}{\textcolor{black}}

\def\edibble{\texttt{edibble}\xspace} \def\sweetpea{Sweetpea\xspace}
\def\tisane{Tisane\xspace} 

\def\dsl{PLanet}

\usepackage{setspace}

\newcolumntype{H}{>{\setbox0=\hbox\bgroup}c<{\egroup}@{}}

\makeatletter
\newcommand\verytiny{\@setfontsize\verytiny{5pt}{6pt}} 
\makeatother

\def\<#1>{\codeid{#1}}
\newcommand{\codeid}[1]{\ifmmode{\mbox{\small\ttfamily{#1}}}\else{\small\ttfamily #1}\fi}
\newcommand{\codeidsmall}[1]{\ifmmode{\mbox{\smaller\ttfamily{#1}}}\else{\smaller\ttfamily #1}\fi}

\lstdefinestyle{planetstyle}{
    backgroundcolor=\color{white},            
    commentstyle=\color{codegreen}\itshape,    
    keywordstyle=\color{magenta},              
    numberstyle=\tiny\color{codegray},        
    stringstyle=\color{codepurple},            
    basicstyle=\ttfamily\small,         
    breakatwhitespace=false,                  
    breaklines=true,                          
    captionpos=b,                             
    keepspaces=true,                          
    numbers=left,                             
    numbersep=5pt,                            
    showspaces=false,                         
    showstringspaces=false,                   
    showtabs=false,                           
    tabsize=4,                                
    alsoletter={/, <-},                       
    deletekeywords={library, cm, order, list, /, <-, count, options}, 
    morekeywords={Units, Design, assign, ExperimentVariable, nest, cross,
    within_subjects, between_subjects, counterbalance, order, num_trials}, 
    literate={/}{{/}}1 {<-}{{<-}}2            
}

\usepackage{booktabs}

\definecolor{ramblerblue}{RGB}{219,234,254}   
\definecolor{ramblerbluetxt}{RGB}{30,64,175}  
\definecolor{baselineamber}{RGB}{254,243,199} 
\definecolor{baselineambertxt}{RGB}{146,64,14}

\lstdefinestyle{edibblestyle}{
    backgroundcolor=\color{white},
    commentstyle=\color{codegreen}\itshape,
    keywordstyle=\color{magenta},
    frame=single,
    numbers=none,
    xleftmargin=0pt,
    framexleftmargin=0pt,
    numberstyle=\tiny\color{codegray},
    stringstyle=\color{codepurple},
    basicstyle=\ttfamily\scriptsize,
    breakatwhitespace=false,
    breaklines=true,
    captionpos=b,
    keepspaces=true,
    numbers=left,
    numbersep=5pt,
    showspaces=false,
    showstringspaces=false,
    showtabs=false,
    tabsize=4,
    alsoletter={\_}, 
    morekeywords={
        design, set_units, set_trts, allot_trts, assign_trts,
        serve_table, crossed, crossed_by, c
    },
    moredelim=[s][\color{codepurple}]{"}{"},
    moredelim=[s][\color{codepurple}]{'}{'}
}

\lstdefinestyle{touchstonestyle}{
    backgroundcolor=\color{white},            
    commentstyle=\color{codegreen}\itshape,    
    keywordstyle=\color{magenta},              
    numberstyle=\tiny\color{codegray},        
    stringstyle=\color{codepurple},           
    basicstyle=\ttfamily\scriptsize,         
    breakatwhitespace=false,                  
    breaklines=true,                          
    captionpos=b,                             
    keepspaces=true,                          
    numbers=left,                             
    numbersep=5pt,                            
    showspaces=false,                         
    showstringspaces=false,                   
    showtabs=false,                           
    tabsize=4,                                
    alsoletter={/, <-},                       
    deletekeywords={library, cm, order, list, /, <-, count, options}, 
    morekeywords={Latin},
    morekeywords=[2]{artist,coffee, baseline, room},
    keywordstyle=[2]\color{codepurple},     
    literate={/}{{/}}1
    {<-}{{<-}}2{\<}{{\textless}}1{\>}{{\textgreater}}1{\{}{{\{}}1{\}}{{\}}}1{,}{{,}}1
}

  \usepackage{booktabs}
  \usepackage{adjustbox}
  \usepackage{tabularx}
  \usepackage{supertabular}
  \usepackage{booktabs}
  \renewcommand{\arraystretch}{1.3}
  \usepackage{array}
  \newcolumntype{L}[1]{>{\raggedright\arraybackslash}p{#1}}
  \newcolumntype{C}[1]{>{\centering\arraybackslash}p{#1}}
  \usepackage{pifont} 
  \def\edibble{\textsc{edibble}\xspace} 
  \def\touchstone{\textsc{Touchstone}\xspace} 
   
  \usepackage{enumitem}

\usepackage{wasysym}  
\usepackage{booktabs} 

\begin{document}

\setlength{\parskip}{0pt}
\title{\dsl: Formalizing and Analyzing Assignment Procedures in the Design of Experiments}

\author{London Bielicke}
\email{londonbielicke@ucla.edu}
\orcid{1234-5678-9012}
\affiliation{%
  \institution{UCLA}
  \city{Los Angeles, CA}
  \country{USA}
}

\author{Anna Zhang}
\email{azhang03@mit.edu}
\orcid{1234-5678-9012}
\affiliation{%
  \institution{MIT}
  \city{Cambridge, Massachusetts}
  \country{USA}
}

\author{Shruti Tyagi}
\email{shrutityagi@ucla.edu}
\orcid{1234-5678-9012}
\affiliation{%
\institution{UCLA}
\city{Los Angeles, CA}
\country{USA}
}

\author{Emery Berger}
\email{emery@cs.umass.edu }
\orcid{1234-5678-9012}
\affiliation{%
  \institution{University of Massachusetts, Amherst}
  \city{Amherst, MA}
  \country{USA}
}

\author{Adam Chlipala}
\email{adamc@csail.mit.edu}
\orcid{1234-5678-9012}
\affiliation{%
  \institution{MIT}
  \city{Cambridge, Massachusetts}
  \country{USA}
}

\author{Eunice Jun}
\email{emjun@ucla.edu}
\orcid{1234-5678-9012}
\affiliation{%
  \institution{UCLA}
  \city{Los Angeles, CA}
  \country{USA}
}

\renewcommand{\shortauthors}{Bielicke et al.}


\begin{abstract}
  Experimental designs reflect assumptions about variable relationships that
  determine what causal queries researchers can answer through the experiment.
  Accounting for and communicating these assumptions is essential for drawing
  valid, generalizable conclusions from scientific experiments. Unfortunately,
  existing experimental design tools elide these details, expecting researchers
  to reason about design decisions and assumptions on their own. To surface
  assumptions and enable design exploration, we introduce a grammar of
  composable operators for constructing experimental assignment procedures
  grounded in matrix algebra. The \dsl{} DSL implements this grammar and
  compiles \dsl{} programs into constraint satisfaction problems over matrices.
  Together, \dsl{}'s composable grammar and matrix representation enable a
  static analysis to determine which causal queries are testable under different
  assumptions. In an expressivity evaluation, \dsl{} was the most expressive of
  existing DSLs. Critical reflections
  with the authors of these DSLs revealed that \dsl{} makes design choices
  explicit without requiring procedural specification. Think-aloud studies
  showed that \dsl{} facilitated design exploration and surfaced assumptions
  researchers may otherwise overlook.  
\end{abstract}



\keywords{experimental design, domain-specific language}


\maketitle

\def\edibble{\texttt{edibble}\xspace} \def\sweetpea{Sweetpea\xspace}
\def\tisane{Tisane\xspace}

\newcommand{\ramblerPill}{\colorbox{ramblerblue}{%
  \textcolor{ramblerbluetxt}{\texttt{Graphical Interface}}}}
\newcommand{\baselinePill}{\colorbox{baselineamber}{%
  \textcolor{baselineambertxt}{\texttt{Baseline}}}}

\section{Introduction}

Experiments are a primary tool in the scientist's toolbox for understanding the world.
Experiments enable scientists to isolate and test causal relationships among
variables of interest. Evidence from experiments can challenge assumptions, test
hypotheses, and inform the development of new theories in a range of scientific
disciplines~\cite{fisher1935designExperiments}. For instance, in human-computer
interaction (HCI), experiments are commonly used to evaluate the efficacy of new
interfaces, understand user behavior, and explore the impact of design choices
on user experience.

Consider the following familiar scenario: 

\textit{A researcher wants to establish whether a new graphical interface leads
to faster task completion than a text-based alternative. To answer this
question, the researcher must design an experiment that isolates the effect of
the interface from other factors that could influence completion time, such as
task type, participant experience, and the order in which conditions are
experienced. The researcher decides to use a Latin square because they have seen
it used in a previously published study. They conduct the study and find that
participants who used the graphical interface completed the task in less time,
concluding that the graphical interface causes people to complete the task
faster than the text-based interface.}

What the researcher did not realize is that they made the implicit assumption
that the interface will have the same effect on completion times no matter the
type of task. If the researcher includes an additional task variable, they must
make additional considerations. For example, should they counterbalance both the
task and interface conditions? Should participants experience all combinations
of task and interface conditions? These subtle considerations reflect
assumptions about whether the effect of the interface depends on task or the
order in which each interface is administered.


This example illustrates a fundamental yet often overlooked tension
when designing experiments. Researchers face practical constraints, most often
time to run an experiment or access to participants, that may require them to
consider alternatives to an ideal fully randomized experiment. These
alternatives embed assumptions about variable interactions and time-based
effects. When these assumptions do not align with the researchers' understanding
of their domain, the resulting design cannot test the effects that might matter
to them or their discipline. Yet, current experimental-design workflows require
the researcher to be aware of a design’s implicit assumptions and manually
ensure that their domain assumptions and causal queries align. As the
number of variables and potential interactions grows, manually reasoning about
which designs preserve which testable effects becomes intractable. For these
reasons, researchers may default to familiar designs, falsely assuming that they
serve their purposes. What is missing is a way to make design decisions,
assumptions, and their connections explicit.

Some existing tools for experimental design constrain researchers to predefined
designs~\cite{eiselmayer2019touchstone2}, while others offer fine-grained
control but require low-level specification~\cite{edibble,
sweetpea}. All these tools lack support for detecting which causal effects the
experiment can test for. Tools that do check for threats to validity target
large-scale online experiments~\cite{tosch2019planalyzer, bakshy2014planout},
where design choices differ from those of human-subjects research. A critical
gap in the landscape of experimental-design tooling is enabling systematic
consideration and comparison of alternative designs suitable for
researchers' goals and assumptions.

We address this gap by developing
 a grammar for expressing assignment
procedures in experimental design. 
Through an iterative design process informed by experimental-design theory and
practice, we identified a set of primitive operations that correspond to matrix
manipulations. 
A key technical insight is that each primitive operation
translates into a logical constraint on an underlying matrix representation. 
We instantiate this grammar in the \dsl{} domain-specific language (DSL) and
implement a static analysis to detect 
when a causal effect is testable.

We assess the expressivity of the \dsl{} DSL compared to \edibble{} and
\touchstone2{}, two existing DSLs for experimental design.
We found that \dsl{} was the most expressive, covering 14 of 15 experiments. In
addition, we engaged the authors of these DSLs in
critical-reflection~\cite{CriticalReflection2019} sessions to dig into how the
design and analysis capabilities of \dsl{} could enhance scientific practice. We
found that \dsl{} provides a language for researchers to articulate design
choices without selecting pre-defined categories or requiring procedural
details, which can facilitate scientific transparency and deepen understandings
of design decisions. 

Finally, in order to observe in-situ how \dsl{} influences researchers'
reasoning about and exploration of experimental designs,
we conducted think-aloud studies with six researchers\footnote{We also
conducted three case studies with researchers using an earlier version of
\dsl{}, which we report in supplemental material and the appendix.} who specified \dsl{} programs
using \dsl{}'s web interface. All participants reported prior exposure to
designs expressible in \dsl{}, but they lacked precise language to describe
them. Yet, after initial exploration with the tool, they could articulate how
design choices reflected different assumptions and hypotheses.


This paper makes the following contributions:
\begin{itemize}[nosep,topsep=4pt,leftmargin=*]
  \item a grammar and DSL defining assignment procedures
  grounded in matrix algebra (\autoref{sec:grammar}, \autoref{sec:dsl});
  \item a static analysis over \dsl{} programs to detect differences in testable
  causal queries between experimental designs (\autoref{sec:analysis});
  \item an evaluation demonstrating \dsl{} DSL's comparative expressivity for a
  broad range of designs found in a sample of recent HCI publications
  (\autoref{sec:evaluation});
  \item critical reflections with experimental-design DSL developers, illuminating how
  \dsl{} DSL enhances scientific practice, communication, and understanding
  (\autoref{sec:expert-eval}); and 
  \item empirical evidence from think-aloud studies with researchers
  demonstrating how \dsl{} facilitates exploration and reasoning about a variety
  of experimental designs (\autoref{sec:user-study}).
\end{itemize}

\section{Background and Related
Work}\label{sec:related-work} We situate our work in the context of
experimental-design theory, tools for designing studies, and domain-specific
languages (DSLs). We define terminology related to experimental design in
\autoref{box:definitions}.

\subsection{Experimental-Design Theory}\label{subsec:design-theory} \keyidea{The
goal of an experiment is to estimate the causal effect of an intervention on an
outcome.} Determining valid causal effects requires that each participant's
outcome is independent of another participant's assignment (i.e.,
non-interference), no third variable influences both the treatment and outcome
variable (i.e., no confounding), and that every participant has a non-zero
probability of experiencing each condition of the treatment variable (i.e.,
positivity). Ideal experiments minimize these threats by randomly assigning
exactly one condition to participants but require large sample sizes. Many
human-subjects experiments therefore use different assignment procedures (e.g.,
within-subjects designs) that require stronger assumptions. \dsl{} helps
researchers surface and compare these assumptions, which are often left
implicit.

\begin{infobox*}
    \begin{tcolorbox}[colback=gray!10!white,colframe=gray!60!black, arc=1mm, top=1mm, bottom=1mm, left=1mm, right=1mm]
    \fontsize{8}{9.5}\selectfont

    
    \textbf{Randomized Design.}
    A design where experimental units are assigned to different conditions
    (groups) at random~\cite{apaDictionaryPsychology}.
    
    \textbf{Between-Subjects Variable.}
    A variable (i.e., factor) for which each subject experiences only one of its
    conditions~\cite{seltman2018experimental}.
    
    \textbf{Within-Subjects Variable.}
    A variable for which each unit experiences more than one
    condition~\cite{seltman2018experimental}.
    
    \textbf{Counterbalancing.}
    Across experimental units, each condition occurs an equal number of times in
    each position of the assigned orders. This technique is used in
    within-subjects designs to mitigate the possible influence of condition
    order on outcomes~\cite{apaDictionaryPsychology}.

    \textbf{Full Counterbalancing.}
    A canonical within-subjects design where every
     order of conditions occurs an equal number of
    times across participants~\cite{apaDictionaryPsychology}.
    
    \textbf{Latin Square.}
    A canonical within-subjects design where each treatment occurs once in
    each position across all orders~\cite{apaDictionaryPsychology}.
    
    \end{tcolorbox}%
    \vspace{-1em}
    \caption{\textbf{Key experimental design terminology.}}
    \label{box:definitions}
    \end{infobox*}

Fisher's \textit{The Design of Experiments} established randomization,
replication, and control as foundational principles for assessing causal
relationships~\cite{fisher1935designExperiments}. Building on this, Campbell and
Stanley~\cite{campbell1963experimental} and Cook and
Campbell~\cite{cook1979quasiExperimentation} decompose experimental rigor into
three components: \textit{assignment, sampling, and measurement}. 
We focus on assignment here and plan to address sampling and measurement in the
future (\autoref{sec:future-work}).

The statistical literature connects experimental design to linear algebra: Cox
and Reid~\cite{CoxReid2000} reason about design matrices in terms of balance and
orthogonality, and Kronecker product methods formalize partial
counterbalancing~\cite{Vartak1955Kronecker}. We ground \dsl{} in these
linear-algebra principles, using matrices as the underlying representation for
experimental designs (\autoref{sec:dsl}).
\subsection{Tools for Designing Studies}
Several DSLs~\cite{gosset,bakshy2014planout}, software
packages~\cite{edibble,blair2019declaring}, and standalone
applications~\cite{mackay2007touchstone,eiselmayer2019touchstone2} specialize in
experiment design. Touchstone2~\cite{eiselmayer2019touchstone2} provides the
\touchstone{} Language (TSL) and an interface for designing within-subjects
experiments based on statistical power. However, its power analysis does not
account for how different assignment procedures may require different
assumptions. In contrast, \dsl{} focuses on helping researchers explore and
compare designs based on different research questions and assumptions.
\edibble~\cite{edibble} is a ``grammar of study design'' that constructs
experiments through hierarchical unit structures. In comparison, \dsl{}
decouples unit specification from condition assignment, enabling users to define
designs by constraining the space of viable plans. Unlike \edibble{}, \dsl{}
also enables researchers to analyze and compare designs
(\autoref{sec:evaluation}).
\sweetpea~\cite{sweetpea} focuses on constructing valid conditions in
multifactorial designs by identifying derived factors. \dsl{} focuses on mapping
conditions to experimental units; these capabilities are orthogonal.
\tisane~\cite{jun2022tisane} specifies conceptual and data-measurement
relationships to infer statistical models but does not capture how conditions
are assigned or distributed across units.
Importantly, none of these DSLs provide analyses to compare experimental designs
or data-collection methods.

PlanOut~\cite{bakshy2014planout} is a DSL for defining large-scale online field
experiments by specifying variables and the number of users. PlanOut supports
multifactorial experiments but does not allow researchers to specify
counterbalanced designs. PlanAlyzer~\cite{tosch2019planalyzer} detects
internal-validity issues for experiments expressed in PlanOut, focusing on
implementation errors such as failures to randomize due to hashing functions.
While PlanAlyzer checks for positivity violations, its analysis is limited to
designs expressible in PlanOut. For example, PlanAlyzer does not detect threats
caused by time-varying effects in within-subjects designs. In contrast, \dsl{}
focuses on structural properties of experimental designs common in controlled
human-subjects experiments.

\subsection{Domain-Specific Languages in HCI}
HCI researchers have developed DSLs for data-related tasks, including data
visualization (e.g., Vega-Lite~\cite{satyanarayan2017vega}),
analytical-hypothesis expression~\cite{suh2022grammarHypo}, and statistical
analysis~\cite{jun2019tea,jun2022tisane,jun2024rtisane}. These serve both
programmers and users of interactive systems built on top of them. Wilkinson's
Grammar of Graphics~\cite{GrammarOfGraphics} is a key precedent: by exposing
primitives for graphics, it made statistical visualizations compositional and
communicable. Similarly, \dsl{}'s grammar exposes primitives that make
experimental designs compositional, communicable, and
analyzable~\cite{blackwell2001cognitive}. We view \dsl{} as one possible
instantiation of this grammar, prioritizing conceptual clarity over
specification efficiency.

\section{Experimental Assignment Grammar} \label{sec:grammar}
Through an iterative design process grounded in examinations of both classical
experimental design theory (\autoref{subsec:design-theory}) and example
human-subjects papers, we identified a set of composable nouns and verbs to
describe experimental assignment procedures in human-subjects experiments.  
\autoref{fig:grammar} summarizes our grammar. 


\subsection{Concepts}~\label{sec:concepts} \keyidea{There are six core concepts (i.e.,
nouns) involved in an experimental assignment.}
 
\noindent \defterm{Units.} The unit is an entity the experimenter wants to observe and
make inferences about. Often, units are individual participants.

\noindent \defterm{Condition.} An experimental condition is a possible assignment value to
a set of experimental variables. Units experience exactly one experimental
condition at a given point in time. 

\noindent \defterm{Trial.}
A trial is an isolated instance of an experimental condition at
a given point in time. Units complete one or more trials. 

\noindent \defterm{Variables.}
Experimental variables are the independent variables the experimenter
manipulates or controls in a study. In the context of an experiment, a variable
is manipulated to be either \textit{between-subjects} or
\textit{within-subjects}. 
An experimental variable consists of one or more conditions. 

\noindent \defterm{Experimental Plan.}
An experimental plan is an order of conditions to which a unit can be assigned. 

\noindent \defterm{Design.} A design specifies \textit{how} the possible plans
are constructed and assigned to units. An empty design contains no variables. 

\subsection{Operators}


A data structure representing experimental plans must be able to represent logical
relationships between plans. For example, experimental designs often require
that all conditions occur an equal number of times across plans. Therefore, we
adopt a matrix representation. 
We define the following operators as constraints over a matrix of plans. 


\defterm{Counterbalance.}
Counterbalancing is a
requirement, or constraint, that each condition of an experimental variable is
observed an equal number of times for each trial number across all plans. 
It accounts for some order effects (i.e., practice, fatigue) in
experiments ~\cite{apaDictionaryPsychology}. 

\begin{figure}[t]
  \centering

    \includegraphics[width=0.8\linewidth]{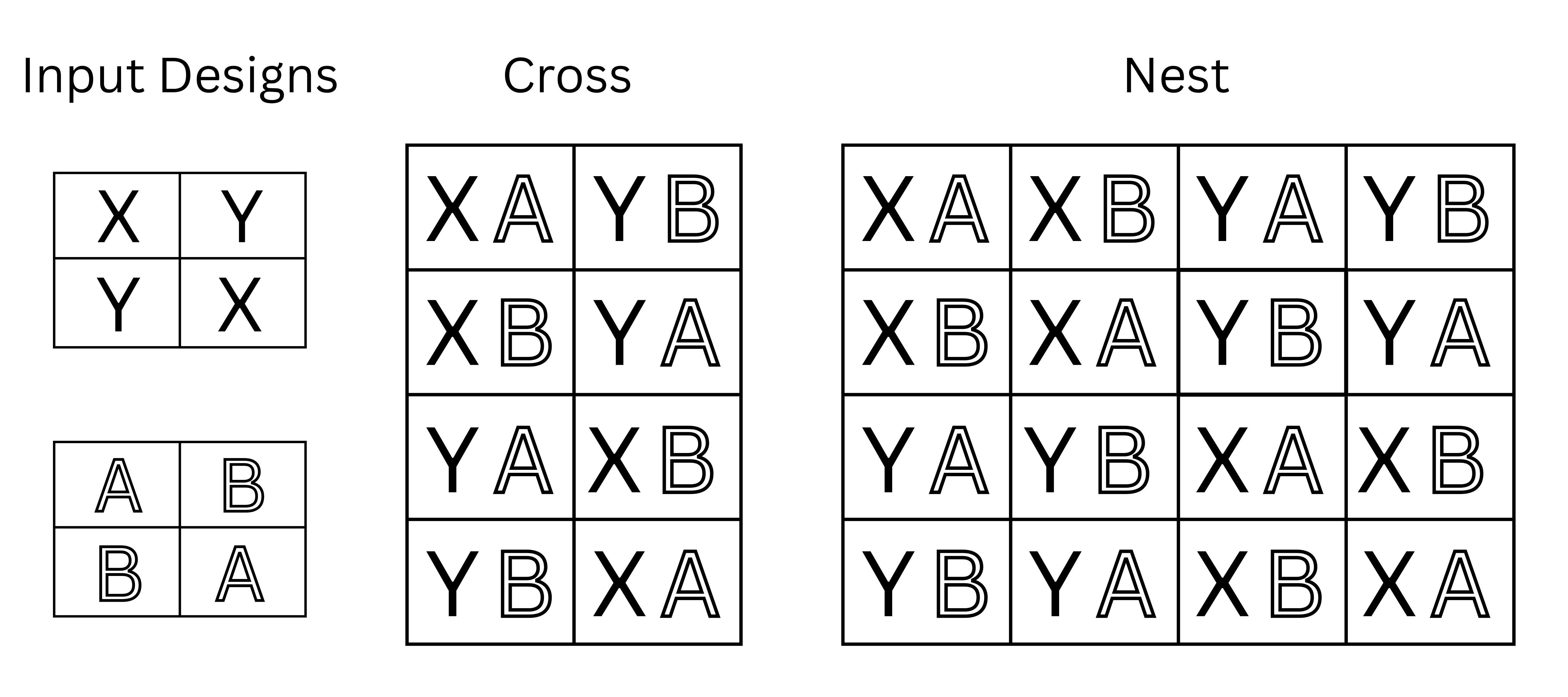}
  
    \caption{\textbf{Composing designs in \dsl{} using \texttt{cross} and
    \texttt{nest}.} In the crossed design (left), every row contains every
    condition of each variable but not every combination (e.g., X and Y appear
    with A and B, but not all combinations XA, XB, YA, YB appear in one row).
    In the nested design (right), the outer condition (A or B) is held fixed
    within each $2\times2$ block while the inner conditions (X and Y)
    alternate.}
    \label{fig:composition}
\end{figure}

\defterm{Cross.}
Cross superimposes every plan in two designs
(\autoref{fig:composition}). Each row of the resulting design matrix is formed by
taking the element-wise pairing of a row from each input matrix. As a result,
the matrix has $n \times m$ rows, where $n$ and $m$ are the number of rows in
the first and second designs, respectively.


\defterm{Nest.}
Nest combines two designs such that the conditions of variables in one design
are held fixed across all conditions of the other design (\autoref{fig:composition}).
Nest maps directly to the \textit{Kronecker product} (defined in \autoref{sec:appendix}), where the element-wise
multiplication corresponds to the combination of conditions. In \dsl{}, users
repeat conditions within a participant by nesting a nonempty design with an
\textit{empty} design.\footnote{The number of trials specified in the
\textit{empty} design determines the number of repetitions. There are two
primary repetition strategies: (i) for each participant, conditions repeat $n$
times in a row, or (ii) for each participant, condition orders repeat $n$ times.
Swapping the order of arguments in \texttt{nest} determines the repetition
strategy.
} 


\defterm{Order.}
Specifying an order enforces that all participants experience conditions of the corresponding variable in the same order. 

\defterm{Limit plans.}
Researchers often restrict the number of possible plans in an experiment. In
\dsl{}, limiting the number of plans sets an upper bound on the height of the
design matrix. For example, a degree-three Latin square is a counterbalanced
design with exactly three plans. Without this restriction, there would be six
possible counterbalanced plans. 

\defterm{Multifact.}
Combines every condition of all input variables to create one multi-factor
variable, which corresponds to taking the Cartesian product of each variable's
conditions. For example, given two variables with two conditions each (1 or 2
and a or b), the four possible conditions are 1-a, 1-b, 2-a, 2-b. 

\begin{figure}[th]
  \centering
  \small
  \begin{align*}
    \text{AssignExp} &\rightarrow \text{assign(UnitExp, DesignExp)} \\
    \text{VariableExp} &\rightarrow \text{Variable(str, str[])} \\
    &\mid \text{multifact(VariableExp[])} \\
    \text{UnitExp} &\rightarrow \text{Units(int)} \\
    \text{DesignExp} &\rightarrow \text{Design()} \\
    &\mid \text{DesignExp.DesignMethod} \\
    &\mid \text{nest(DesignExp, DesignExp)} \\
    &\mid \text{cross(DesignExp, DesignExp)} \\
    \text{DesignMethod} &\rightarrow \text{counterbalance(VariableExp)} \\
    &\mid \text{between\_subjects(VariableExp)} \\
    &\mid \text{within\_subjects(VariableExp)} \\
    &\mid \text{limit\_plans(int)} \\
    &\mid \text{num\_trials(int)} \\
    &\mid \text{order(VariableExp, str[])}
  \end{align*}
  \caption{\textbf{Our formal grammar of experimental assignment.}}
  \label{fig:grammar}
\end{figure}

\newcommand{\cross}{\texttt{cross}\xspace}
\newcommand{\nest}{\texttt{nest}\xspace}

\newcommand{\PLanetProgram}{
\begin{figure}[t]
    \centering
    \begin{subfigure}[t]{0.65\textwidth}
        \centering
        \lstset{
            basicstyle=\ttfamily\scriptsize,
            frame=single,
            language=Python,
            numbers=none,
            xleftmargin=0pt,
            framexleftmargin=0pt,
        }
        \lstinputlisting[ language={Python}, style=planetstyle,   basicstyle=\ttfamily\footnotesize]{example.py}
    \end{subfigure}

\caption{\textbf{Sample
\dsl{} program (left) to construct the design described in
\textit{FFL: A Language and Live Runtime for Styling and Labeling Typeset Math Formulas}~\cite{wu2023ffl}}.  
Participants always see creation tasks before
editing tasks. Task number and interface are crossed (line 35). Each participant sees every task number and every interface
but not every combination of task number and interface. This rule holds 
for both task types: creation and editing. As a result, the trials with task 
number and interface are nested in the task type (line 36). \dsl{} is implemented in Python with a fluent
interface style.}
\label{lst:PLanetProgram}
\end{figure}

}

\newcommand{\PLanetUI}{
\begin{figure*}[t]
    \centering
    \includegraphics[width=.9\textwidth]{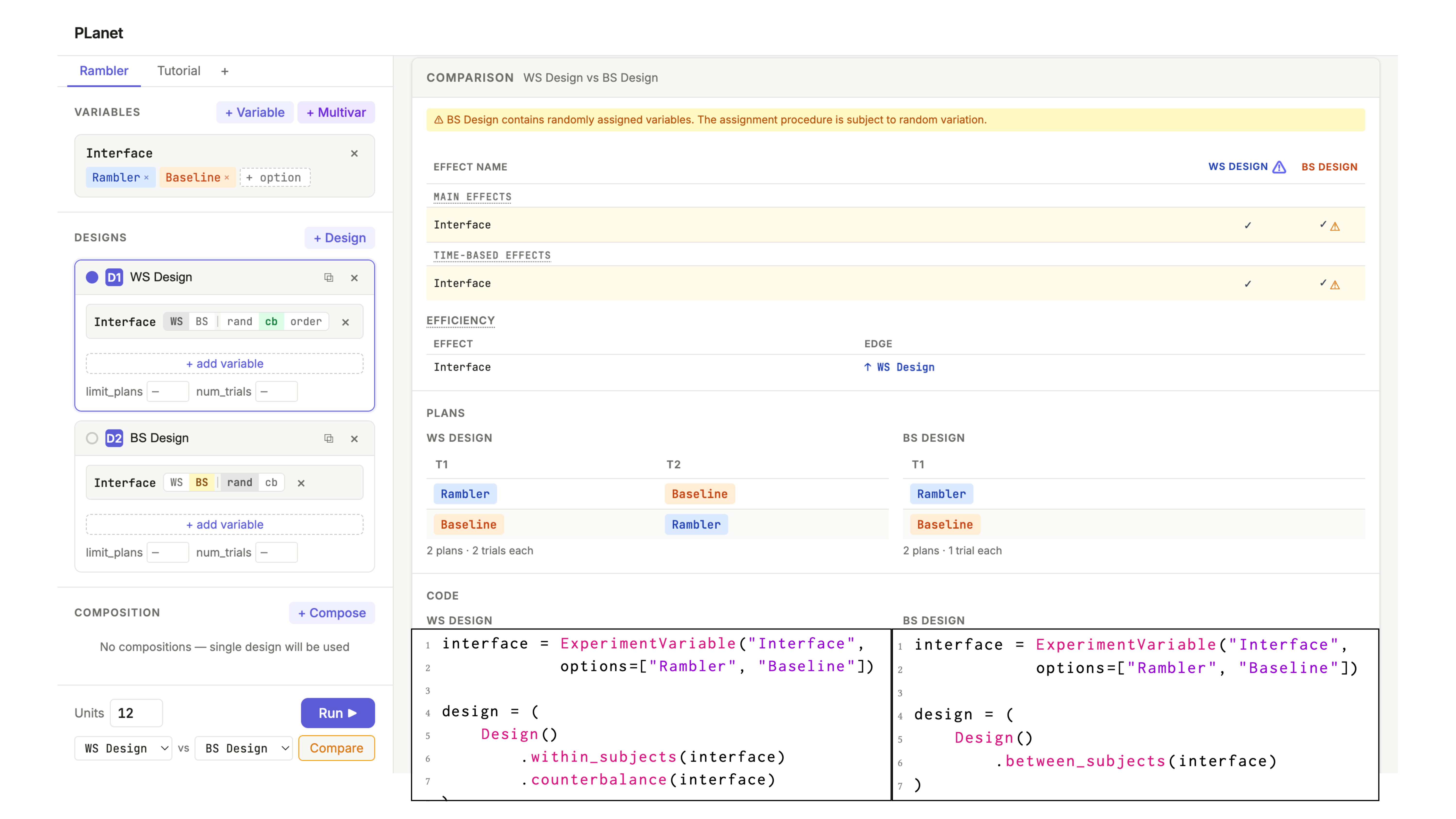}
    \caption{\textbf{\dsl{}'s user interface comparing two experimental designs from our user
    evaluation (\autoref{sec:user-study}).} The interface compiles user
    specifications (left panel) into \dsl{} programs, as shown beneath the interface. The design
    on the left is counterbalanced and within-subjects. The design on the right
    is randomized and between subjects. The comparative analysis identifies the
    within-subjects design as more efficient and warns that it is subject to
    carry-over effects (upper right corner).}
    \label{fig:ui}
\end{figure*}
}

\newcommand{\PLanetProgramDoubleColumn}{
\begin{figure}[!htbp]
    \centering
    \begin{minipage}[c]{0.5\columnwidth}
        \lstinputlisting[language={Python}, style=planetstyle, basicstyle=\ttfamily\fontsize{6}{6}\selectfont, breaklines=true, breakatwhitespace=true]{example.py}
    \end{minipage}
    \hfill
    \begin{minipage}[c]{0.48\columnwidth}
        \includegraphics[width=\linewidth]{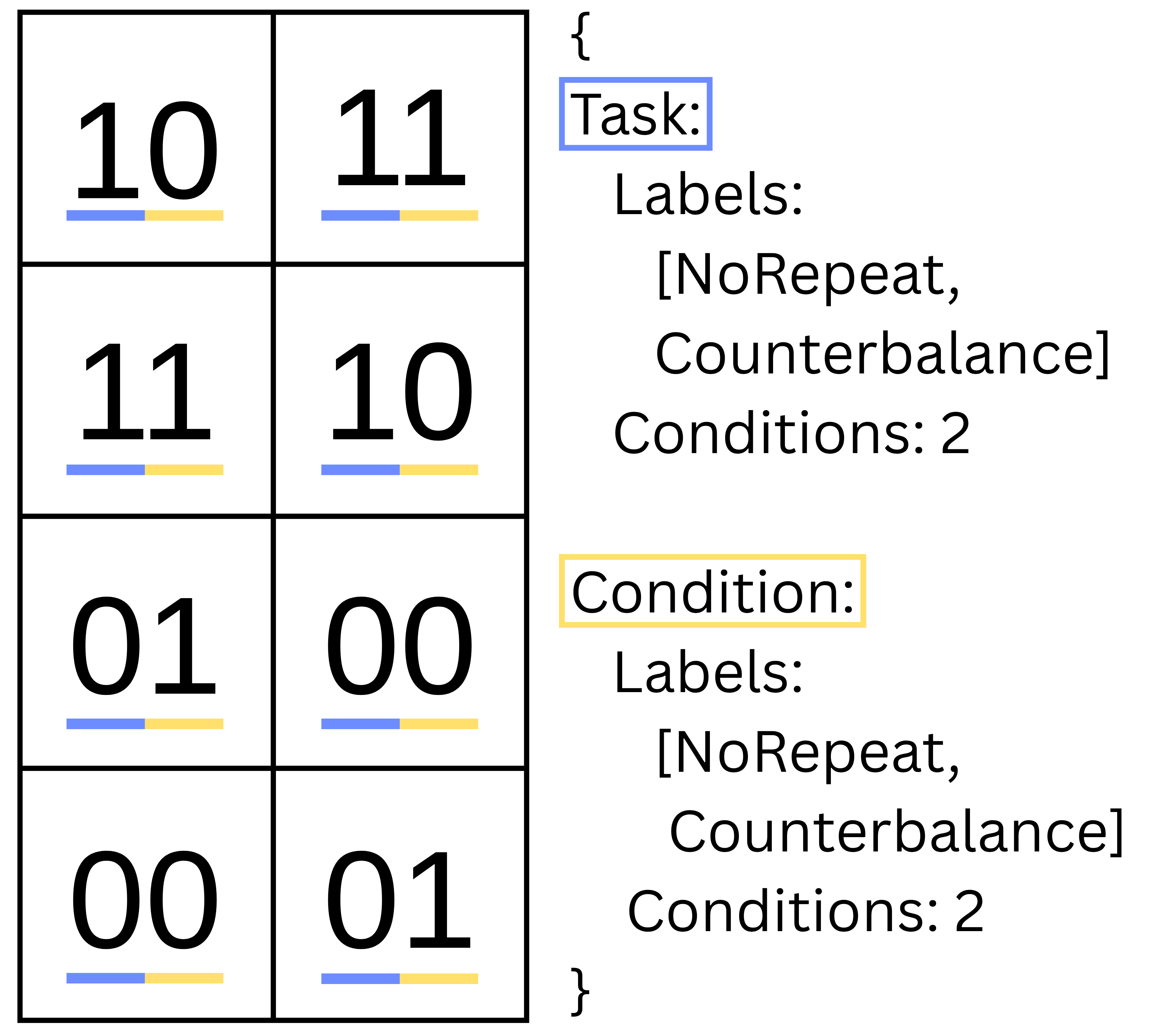}
    \end{minipage}
    \caption{\textbf{\dsl{} program (left) representing an experiment included
     in our expressivity evaluation~\cite{WuARTiST2024}
     (\autoref{sec:evaluation}) and its corresponding matrix representation
     (right).} Task and condition are crossed (line 11). Each participant sees
     every task and every condition but not every combination of the two. The
     task and condition variables correspond to the left and rightmost bits,
     respectively. \dsl{} uses the variable labels to perform static analysis
     checks.}
    \label{fig:PLanetProgram}
\end{figure}
}

\section{\dsl{} Implementation} \label{sec:dsl} 

  


We instantiate this grammar as the \dsl{} DSL.
The DSL then serves as the basis for the graphical interface (\autoref{fig:ui}).
Below, we provide an overview of \dsl{} DSL's implementation and explain how the
interface options map to DSL operations.

\subsection{Generating Plausible Experimental
Plans}~\label{subsec:generate-plans} 
\begin{figure}[h]
    \centering
    \includegraphics[width=\columnwidth]{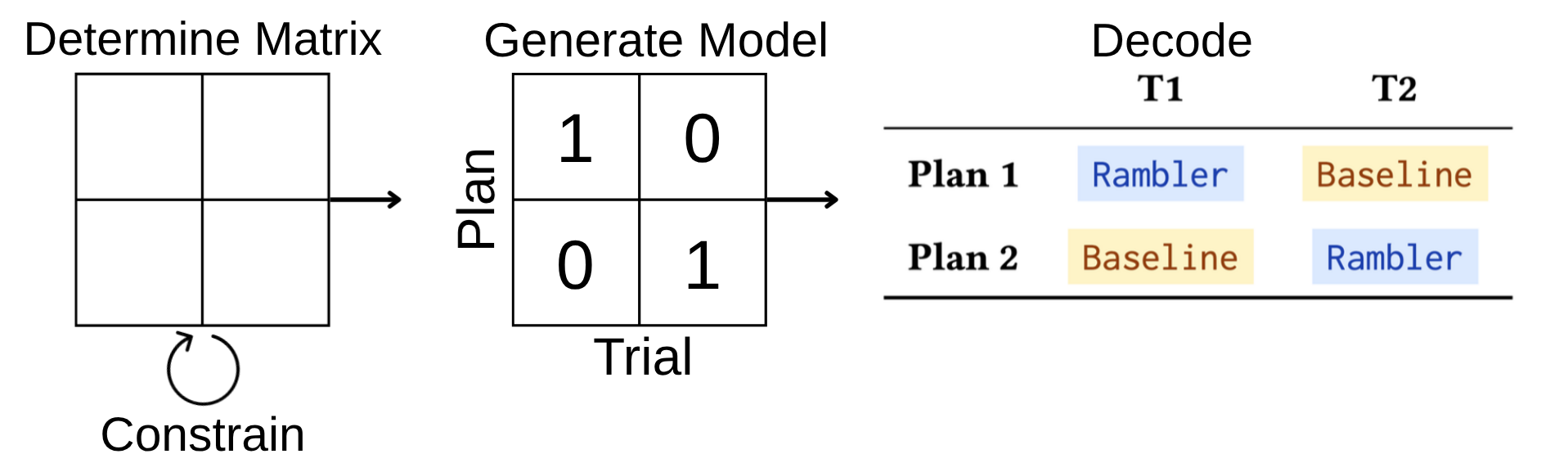}
    \caption{\textbf{Generating viable experimental plans. } 
    \dsl{}
    determines the shape of the design matrix and places constraints
    on entries of the matrix (left) before generating a Z3 model (middle). 
    The numbers in the matrix map directly to
    specific values of a bitvector encoding, which represents possible
    assignments to a set of variables. The last step translates the matrix to a
    table with all viable experimental plans (right). 
    }
    \label{fig:plans}
  \end{figure}

\keyidea{\dsl{} formulates experimental-plan generation as a
constraint-satisfaction problem over matrix entries.} Researchers specify design
requirements, such as counterbalancing, which translate into first-order logic
constraints over matrix entries.  If
the number of units is insufficient to satisfy certain properties, \dsl{} will
throw an error and suggest the minimum number of participants required to
satisfy their design specification. 
The result of successfully running a \dsl{} program is a
table of plans. \autoref{fig:plans} illustrates the process. 

\keyidea{There are three technical challenges to generate valid assignments:}
(i) defining the core operations as logical constraints over matrices, (ii)
solving these constraints efficiently to generate experiment plans, and (iii)
reasoning over both individual experimental variables and their combinations.

\subsubsection{Defining operators as logical expressions.}
A key insight is that \texttt{counterbalance} is
an essential operator in human-subjects experiments.
\dsl{} formalizes counterbalancing as a logical constraint that each possible
condition appears the same number of times in each column of the design matrix.
Counterbalancing can be used to specify both fully counterbalanced and Latin
square designs. 


Then, the \cross and \nest operators direct the application of constraints like
counterbalancing to specific elements or submatrices. For instance, \cross
ensures the constraints from two designs apply element-wise in the resulting
matrix. \nest ensures constraints apply to predefined submatrices. 
 \texttt{order} introduces row-wise constraints specifying the order in which
conditions can appear.

\PLanetUI{}

\subsubsection{Solving Constraints.}
\keyidea{The second challenge is to find
values for trials that satisfy user-specified constraints.} 
The constraints researchers can place on experimental plans may be complex and interdependent. 
Therefore, \dsl{} uses an SMT solver to find assignments of conditions to trials that satisfy all constraints.
Satisfiability
Modulo Theories (SMT) solvers are widely used for constraint satisfaction and
program verification. 
For designs that restrict the number of plans (i.e., \texttt{limit\_plans}), the
SMT solver prunes the search space before generating a set of plans.
Because the solver only returns assignments that satisfy all constraints, design
properties hold by construction.
\dsl{} uses Z3\cite{z3de2008} as its SMT solver.

\keyidea{SMT solver performance varies for different design specifications and
is more efficient when solving for designs (e.g., Latin squares) that do not use
all possible permutations.} Designs that restrict the number of plans are common in
human-subjects experiments with small samples of participants. For fully
counterbalanced designs, the solver must explore many equivalent solutions since
they include every possible permutation of conditions. These solutions differ
only in row order, which is later randomized when assigning plans to units. All
programs in our evaluation (\autoref{sec:evaluation}) execute in under 33
seconds. See performance details in \autoref{sec:appendix}.

\subsubsection{Representing Constraints using Bitvectors.}
\keyidea{The third challenge is to solve a system of constraints pertaining to
individual experimental variables and combinations of experimental variables.}
Supporting both cases is necessary to express a wide range of
multivariate experiments. To support reasoning about individual variables in
multivariate experiments, \dsl{} uses bitvectors in Z3 to represent condition
assignments across all variables. Masking specific segments isolates
individual variables or groups of variables. Applying logical constraints to
these masked bitvectors enables flexible reasoning over both single and
multivariate variables.


\subsection{Graphical User Interface}

\keyidea{The \dsl{} GUI allows users to specify, inspect, and
compare designs without writing code directly.} 
The GUI provides a unified 
environment for specifying designs and viewing their corresponding analyses (\autoref{fig:ui}).
Each design configuration in the interface corresponds to a \dsl{} DSL
program. 
The GUI is implemented as a web
application. 




\section{Analyzing assignment procedures} \label{sec:analysis} 

We check assignment
procedures specified in \dsl{} for (i) testable causal effects and (ii) implicit
assumptions. Both are important for assessing the validity of experiments~\cite{shadish2002validity}. 
\dsl{} checks for these statically without instantiating a specific set of
plans.

\subsection{Testable Causal Effects} 
A single experiment can answer multiple research questions, each of which may
require estimation of a different causal effect. 
For example, researchers may run an experiment to test the average causal effect of a new
interface on a sample to then generalize to the entire population. 
Researchers may also want to assess how the effect of a new interface varies
across different tasks (i.e., an interaction effect). 
\dsl{} supports analysis of three effects: 
\begin{itemize}[nosep,leftmargin=*]
    \item A \textit{main effect} is the effect of an independent variable on the dependent
variable, ignoring all other independent variables.
A main effect is testable when the variable has two or more conditions and all conditions appear in the
design. 

    \item An \textit{interaction effect} is the effect of an independent variable that depends on another independent variable. 
Importantly, interaction effects mean the main effect represents an average that
may not generalize to any particular subgroup. Each subgroup may have a different causal effect estimate. 


    \item A \textit{time-based effect} is the effect caused by \textit{when} a
participant takes part in a trial. For example, participants may
complete a task faster later in a study due to \textit{learning
effects}~\cite{apaDictionaryPsychology}. 
Time-based effects can impact interpretation of main and interaction effects. 
\end{itemize}

\subsubsection{Determining if an effect is testable}
An effect is defined to be \textit{testable} if and only if it can be estimated
without bias (i.e., systematic deviation from its true
value)~\cite{pearl2009causality} from data collected according to the assignment
procedure. The presence of bias is detected by checking the structure of the design. 

In experiments, there are three primary threats to testability: (i) unit
interference, (ii) confounding, and (iii)
non-positivity~\cite{HernanWhatIf2025}.  
In within-subjects experiments, participant attributes, the environment,
and time-based effects can confound a causal effect. Noninterference is a
standard assumption in the design of experiments~\cite{Cox1959Experiments}, and
we assume it holds in \dsl{}. Because \dsl{} randomly assigns plans to
participants, we can also safely assume no confounding of participant
attributes. 

\dsl{} checks for time-based confounding and positivity violations. 
\dsl{} checks for positivity of main, interaction, and time-based effects.
Positivity requires that every condition has a nonzero probability of being
assigned~\cite{pearl2009causality}. \dsl{} checks and ensures a stronger
property: that each condition is \textit{guaranteed} to appear at least once.
This stronger guarantee is important for small sample sizes, where a nonzero
assignment probability does not ensure that every condition is actually
observed. 

We implement static analyses to check for time-based confounding and positivity
violations, which we describe below. 

\subsection{Static analysis algorithm}
Users of \dsl{} specify desired properties of their design (e.g.,
counterbalance). These are translated into variable labels.
\dsl{} checks which effects are testable by reasoning over these labels and the
dimensions of the design matrix (\autoref{fig:PLanetProgram}). 

\subsubsection{Checking Main Effects.}~\label{sec:main-effects} \dsl{} checks
that a variable's main effect is testable by verifying three requirements.
First, the variable must have two or more conditions so that there are multiple
groups to compare. Second, the number of trials must at least equal the number
of conditions, ensuring that there are a sufficient number of trials to observe
each condition. Third, no condition can repeat within a participant. When
designs are crossed or nested, \dsl{} checks that the main effect is testable in
at least one sub-design.

\subsubsection{Checking Interaction Effects.}
\dsl{} checks that the interaction effect between variables is testable by
verifying requirements for four subcases.

\PLanetProgramDoubleColumn{}

When there is a \texttt{Multifact} variable, \dsl{} applies the checks for main
 effects to the combined conditions of the multifactorial variable. When
 multiple individual variables are included in the design, \dsl{} checks that at
 least one of two requirements is satisfied. First, both variables are
 counterbalanced and the number of plans is equal to $n! \cdot n!$, where $n$ is the
 number of conditions of each variable. Below this bound, at least one
 combination of conditions may never appear in the design. Second, one variable
 is counterbalanced and the other has a fixed order. The fixed-order variable
 ensures each of its conditions appears in a distinct position. Because the
 conditions of the counterbalanced variable appear in all positions, every
 condition of the counterbalanced variable co-occurs with every condition of the
 fixed-order variable.

When variables are composed using \texttt{nest}, \dsl{} checks that the
corresponding main effects are testable in at least one of its sub-designs. 
 Nesting takes the Kronecker product of two designs (\autoref{sec:dsl}),
guaranteeing that every condition in one sub-design is paired with every
condition in the other.

When variables are composed using \texttt{cross}, \dsl{} checks that at
least one variable is counterbalanced and the main effect of the other
variable is testable in its corresponding sub-design. Crossing takes the
Cartesian product of the set of plans from two designs (\autoref{sec:dsl}),
guaranteeing that every condition in each position of one sub-design is paired
with every condition at the same position in the other design.

\subsubsection{Checking Time-Based Effects.}
\dsl{} checks that a time-based effect is testable by verifying that (i) the
corresponding main or interaction effect of the variable is testable and (ii) the
variable has a \texttt{Counterbalance} constraint so that every condition
appears an equal number of times in each position across all plans. 
%


\subsection{Comparing designs for implicit assumptions}

\dsl{} compares the assumptions about time-based and interaction effects that
two experimental designs make. \dsl{} checks which design requires additional
assumptions to test the true effect of a variable. We chose to focus on
comparing the assumptions between designs rather than enumerating all
assumptions for two reasons: (i) Researchers make numerous assumptions when
designing experiments, many of which are not actionable. For example, a
researcher, in designing an experiment with a set of variables, assumes that the
causal effect of interest does not depend on any other variable
\textit{excluded} from the study (e.g., phase of the moon). 
(ii) We conjecture that comparing how two designs differ helps researchers
contextualize trade-offs or possible revisions to the designs. 

When comparing two designs, \dsl{} determines whether each variable (or variable
combination) is compared within- or between-subjects. This distinction matters
because within-subjects comparisons eliminate between-person variability and,
therefore, require fewer participants to detect an effect~\cite{Charness212Methods}. 
%
Checking whether a variable is compared within-subjects is straightforward for
simple designs but becomes more difficult for designs with (i) multiple
within-subjects variables, (ii) both within-subjects and between-subjects
comparisons (e.g., a mixed design), (iii) crossing, (iv) nesting, or (v) any
combination of the above. 

A variable is compared within-subjects if and only if (i) it has a
\texttt{NoRepeat} constraint and (ii) the number of trials per participants
equals the number of conditions. If both checks pass, this means every
participant is guaranteed to see every condition. When designs are nested,
\dsl{} checks for within-subjects comparisons of combined conditions by
verifying that both properties hold within each subdesign. 

If one design compares a variable within-subjects and the other does not, the
within-subjects design will require fewer participants to detect that variable's
effect, because it eliminates between-person variability. If both designs
compare a variable within-subjects, the design with more trials per participant
will require the same or fewer participants. This variable-level comparison
allows researchers to see the trade-offs between designs without making
assumptions about effect sizes.

\newcommand{\rqPLanetExpressivity}{\textbf{RQ1: \dsl{} Expressivity.}}
\newcommand{\rqExpressivityComparison}{\textbf{RQ2: Comparative Expressivity.}}
\newcommand{\rqDSLDesign}{\textbf{RQ3: DSL Semantics.}}

\newcommand{\Support}{\CIRCLE}
\newcommand{\PartialSupport}{\LEFTcircle}
\newcommand{\NoSupport}{\Circle}

\newcommand{\NestComparisonFigure}{
\begin{figure}[t]
    \captionsetup{labelfont=bf, textfont=normalfont}
    \centering
    \begin{subfigure}[c]{0.49\textwidth}
        \centering
        \lstset{
            basicstyle=\ttfamily\scriptsize,
            frame=single,
            language=Python,
            numbers=none,
            xleftmargin=0pt,
            framexleftmargin=0pt,
        }
        \lstinputlisting[
            style=planetstyle,
            basicstyle=\ttfamily\scriptsize,
        ]{sweating_details.py}

        \caption{\texttt{\dsl{}}}
    \end{subfigure}
    \hfill
    \begin{subfigure}[c]{0.47\textwidth}
        \centering
    \includegraphics[width=\linewidth]{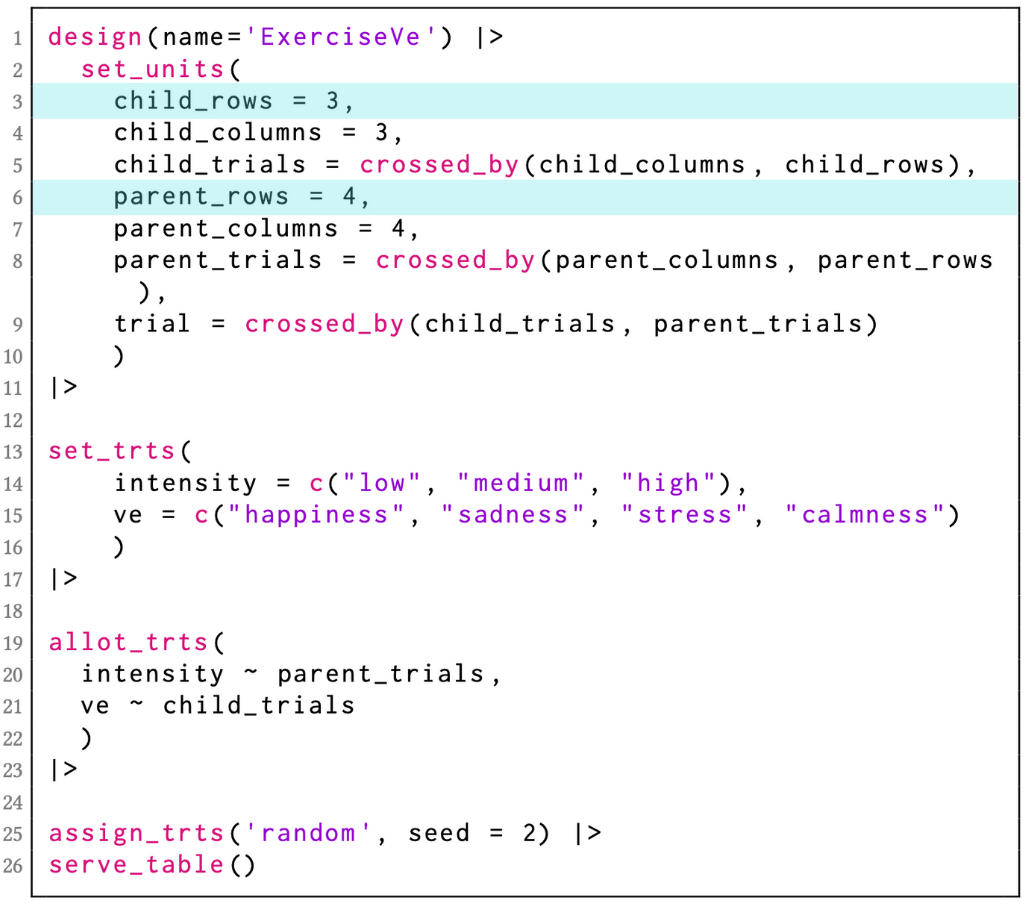}
    \caption{\texttt{\edibble{}}}
    \end{subfigure}
    \caption{\textbf{Nested design from \textit{Sweating the Details: Emotion
    Recognition and the Influence of Physical Exertion in Virtual Reality
    Exergaming} ~\cite{PottsSweatingDetails2024} implemented in \dsl{} and
    \edibble{}}. \dsl{} correctly and explicitly represents that both the
    \texttt{Exercise Intensity} and \texttt{Emotion VE} conditions are
    counterbalanced and that there are 72 participants. The \edibble{} program
    does not correctly represent this design.
    \edibble{}’s strict unit hierarchy splits participants across
    \texttt{child\_row} (b, line 3) and \texttt{parent\_row} (b, line 6). Thus,
    participants do not appear in the trial table, leaving the user to interpret
    the participant-to-condition mapping. Additionally, edibble only allows
    constructing this design with exactly 12 participants, but there were 72
    participants in the original experiment.}

    \label{fig:nest-comparison}
\end{figure}
}

\newcommand{\perPaperResults}{
    \begin{table*}[t]
        \centering
        \caption{\textbf{The 15 sampled experiments and summaries of their assignment
        procedures}. \dsl{} is the most expressive (14 of 15 experiments),
        followed by \edibble{} (12 of 15) and
        \touchstone{} (8 of 15) (\autoref{sec:evaluation}).  \textmd{Circles
        indicate whether the system could express the experiment (\Support\ =
        expressible, \PartialSupport = partially expressible, or \NoSupport\ = not expressible,).}}
        \label{tab:perPaperResults}
        \centering
            \footnotesize
            \begin{tabular}{p{4.5cm} L{2cm} p{4.5cm} p{1cm} p{1cm} p{1cm}}
        \toprule
        PAPER TITLE & KEYWORDS & SUMMARY OF ASSIGNMENT PROCEDURE & \dsl{} & \edibble{} & \touchstone{} \\
        \midrule
        Silver-Tongued and Sundry: Exploring Intersectional Pronouns with
        ChatGPT~\cite{Fuji2024GPT} & User study, Within-subjects & Each
        participant (n=201) experienced 10/11 Japanese pronouns in a random order.
        & \Support & \Support & \Support \\

        When Recommender Systems Snoop into Social Media, Users Trust them Less
        for Health Advice~\cite{Sun2023Recommender} & User study,
        Between-subjects & Each participant (n=341) was randomly assigned one of
        12 conditions. &
        \Support & \Support & \Support \\

        Understanding Perception of Human Augmentation: A Mixed-Method
        Study~\cite{Villa2023Augmentation} & User study, Between-subjects &
        Every participant (n=506) experienced one augmentation type and
        one augmented avatar condition. & \Support & \Support & \Support \\

        Seated-WIP: Enabling Walking-in-Place Locomotion for Stationary Chairs
        in Confined Spaces~\cite{ChanSeatedWIP2024} & User study,
        Counterbalanced, Within-subjects, Latin square & Each participant (n=18)
        is assigned one of nine conditions based on three footstep patterns and
        three posture modes. Footstep and posture counterbalanced via Latin
        square. & \Support & \Support & \Support \\

        Rambler: Supporting Writing With Speech via LLM-Assisted Gist
        Manipulation~\cite{Lin2024rambler} & User study, Counterbalanced,
        Within-subjects & Participants (n=12) experienced both Rambler and baseline in a
        within-subjects, counterbalanced design. & \Support & \Support &
        \Support \\

        Shaping Compliance: Inducing Haptic Illusion of
        Compliance in Different Shapes with Electrotactile
        Grains~\cite{jingu2024shaping} (Exp 2)& User study, Counterbalanced, Within-subjects, Latin square & Each
        participant (n=12) experienced four electrode shapes three times in
        orders determined by a Latin square. & \Support & \Support & \PartialSupport \\
       
        MouseRing: Always-available Touchpad Interaction with IMU
        Rings~\cite{shen2024mousering} (Exp 1)& User study, Within-subjects & Participants (n=12) assigned twelve
       input methods to complete three random tasks, repeated 10 times. & \Support & \Support &
       \PartialSupport \\
     
        MouseRing: Always-available Touchpad Interaction with IMU
        Rings~\cite{shen2024mousering} (Exp 2)& User study, Within-subjects & Participants (n=12) experienced four input
        methods 50 times each. & \Support & \Support & \NoSupport \\

        Skinergy: Machine-Embroidered Silicone-Textile Composites as On-Skin
        Self-Powered Input Sensors~\cite{YuSkinergy2023} & User study,
        Within-subjects & Each participant experienced a random order of eleven
        distinct gestures. & \Support & \Support & \NoSupport \\
        
        ARTiST: Automated Text Simplification for Task Guidance
        in Augmented Reality~\cite{WuARTiST2024} (Exp 1)& User study, Within-subjects,
        Counterbalanced & Each participant (n=16) completed five \textit{unique}
        tasks \textbf{and} five \textit{unique} conditions in five trials. & \Support &
        \Support & \NoSupport \\

        ARTiST: Automated Text Simplification for Task Guidance
        in Augmented Reality~\cite{WuARTiST2024} (Exp 2) & User study, Within-subjects,
        Counterbalanced & Each participant (n=16) performed two unique tasks
        with both ARTiST and baseline
         across two trials. & \Support & \Support & \NoSupport \\

        OK Google, Let's Learn: Using Voice User Interfaces for Informal
        Self-Regulated Learning of Health Topics among Younger and Older
        Adults~\cite{DesaiChin2023} & User study, Counterbalanced,
        Within-subjects & Participants (n=51) are assigned three learning
        strategies. Learning strategy is within-subjects counterbalanced. Task
        is randomly assigned. & \Support & \Support & \NoSupport \\

        Don’t Just Tell Me, Ask Me: AI Systems that Intelligently Frame
        Explanations as Questions Improve Human Logical Discernment Accuracy
        over Causal AI explanations~\cite{Danry2023Explainable} & User study,
        Within-subjects and Between-subjects (mixed) & Each participant (n=204)
        is randomly assigned one of the intervention types and 10/40
        statements. & \Support & \NoSupport & \NoSupport \\

        Sweating the Details: Emotion Recognition and the Influence of Physical
        Exertion in Virtual Reality Exergaming~\cite{PottsSweatingDetails2024} &
        User study, Counterbalanced, Latin square, Within-subjects & Latin-square
        design where participants (n=72) experienced four emotion levels 
        under each of the three intensity levels.  & \Support & \NoSupport & \Support \\

        Shaping Compliance: Inducing Haptic Illusion of
        Compliance in Different Shapes with Electrotactile
        Grains~\cite{jingu2024shaping} (Exp 1)& User study, Counterbalanced, Within-subjects, Latin square & Each
        participant (n=12) experienced twelve conditions based on three grain
        levels and four electrodes determined by a multifactorial Latin square. &
        \NoSupport & \NoSupport & \NoSupport \\

        \bottomrule
        \end{tabular}

    \end{table*}
}

\newcommand{\PropertyComparisonTable}{
\begin{table*}[t]
    \centering
    \caption{\textbf{Comparison of the primary experimental-design properties
    across the three DSLs}. \dsl{} natively supports properties that are
    foundational to HCI experiments.
    Circle
    indicates the type of support (\CIRCLE\ = supported, \LEFTcircle\ = partial
    support, \Circle\ = not supported). The following text denotes how the
    system supports the property. Specific limitations are \textit{italicized}.}
    \label{tab:perPropertyResults}
    \setlength{\tabcolsep}{4pt}
    \renewcommand{\arraystretch}{1.3}
    \footnotesize
    \begin{tabular}{p{2.2cm}|p{3.9cm}|p{3.9cm}|p{3.9cm}}    
    \hline
    \textbf{Design Property} & \textbf{\dsl{}} & \textbf{\edibble{}} & \textbf{\touchstone} \\
    \hline

   \textbf{Latin square} 
   & \Support{}  \texttt{counterbalance} + \texttt{limit\_plans} & 
   \PartialSupport{} Crossing units
   \newline  \textit{Only possible when number of participants is equal to the number of conditions}
   & \Support{} \texttt{Latin} function \\
   \hline

   \textbf{Counterbalancing} 
   & \Support{}  \texttt{counterbalance}
   
   & \PartialSupport{}  Crossing or nesting units
   \newline \textit{Complete randomization and
    counterbalancing are indistinguishable}
   
   & \PartialSupport{}  Predefined functions (e.g., Latin, Full) \newline
   \textit{
   Counterbalancing is not supported outside of predefined functions} \\
   \hline

  \textbf{Full Randomization} 
  & \Support{}  Default behavior  & 
   \NoSupport{}
   \textit{Complete randomization and
   counterbalancing are indistinguishable} &
   \NoSupport{}  
   \textit{All units experience the same order when using the \texttt{Random} function} \\
  \hline

    \textbf{Within-Subjects} 
    & 	\Support{}  \texttt{within\_subjects} operator
    & 	\PartialSupport{} Crossing or nesting units \newline \textit{Participant is not the
    primary unit}
    & \PartialSupport{}  Specific class of functions (i.e.,
    Random, Latin) \newline \textit{Within-subjects designs outside of
    predefined classes are not supported}  \\ \hline

    \textbf{Repetition} 
    & \Support{} Repeating \texttt{ExperimentVariable} options using Python operator
    & \PartialSupport{} Nesting units \newline \textit{Participant is not the
    primary unit}
    & \Support{} \texttt{Block} parameter + \texttt{Serial} function 
    \\
    \hline

  \textbf{Between-Subjects} 
  & 	\Support{} \texttt{between\_subjects} operator
  & 	\Support{}  Defining a single unit
  & 	\Support{}  \texttt{Between} function \\ \hline

  \textbf{Multi-Variable} 
  & \Support{} (i) Composing variable conditions (e.g., \texttt{multifact}) or
  (ii) systematically composing designs (e.g., \texttt{nest}, texttt{cross}) &
  \Support{} \texttt{allottrts} with two or more variables &  \Support{} (i) Composing variable
  conditions using \texttt{cross} or (ii) nesting designs  \\  
 \hline
    
    \end{tabular}
    \end{table*}
}

\section{Expressivity Evaluation}\label{sec:evaluation}

We compared \dsl{} to two DSLs for designing experiments:\\
\edibble{}~\cite{edibble} and \touchstone2{}~\cite{eiselmayer2019touchstone2}.
\touchstone2{} was created by and for the HCI community\footnote{\touchstone2{}
contributes an interactive interface for designing experiments and its
underlying grammar, the \touchstone{} Language (TSL). For the evaluation, we use
TSL because it supports a broader range of experiments than the \touchstone2{}
interface. 
}, while \edibble{} aims to offer a composable 
grammar for defining experimental designs across  
disciplines~\cite{edibble}. 


\subsubsection{Sampling HCI papers and experiments.}
We purposively sampled from the 2023 and 2024 ACM CHI and UIST proceedings 
to identify a rich range of experimental designs. We identified candidate papers
using ``User study'' as a CCS concept or experimental-design keywords in the
body, then sampled two papers from each of six categories (e.g.,
between-subjects only, mixed, counterbalanced, Latin square) to span a wide
range of design complexity. 
In total, our corpus consisted of 12 papers (15 experiments,
as three reported multiple experiments~\cite{shen2024mousering,
WuARTiST2024}).
The Appendix contains more details about our sample. 

\subsubsection{Evaluating programs for sampled papers.} 
The first author 
implemented each experiment
in the corpus using \dsl{}, \edibble{} and \touchstone{}. They studied
documentation, read associated papers, reviewed source code, and followed
available tutorials to develop a deep understanding of each tool. When
necessary, they contacted the tool creators and paper authors for clarification.
The last author reviewed all programs. We evaluated programs
based on correct output and semantic alignment with DSL operations\footnote{By
semantic alignment, we mean interpreting and using DSL constructs based on their
intended meanings. We do not count papers as expressible if doing so would
require misusing the DSL (i.e., if there is no logical mapping between the DSL constructs
and the structure of the actual experiment).}.

\PropertyComparisonTable{}

\subsection{Findings}
Across the DSLs, \dsl{} was the most expressive, fully expressing 14 out of 15
 experiments. \edibble{} and \touchstone{} could express 12/15 and 8/15
 experiments, respectively. Appendix \autoref{tab:perPaperResults} 
 summarizes the overall expressivity results. Notably, the design that \dsl{}
 could not express was inexpressible in both \edibble{} and \touchstone{}. The
 15 programs used every operator except for \texttt{order}.

Similar to \dsl{}, \edibble{} aligns with established experimental-design
theory. However, despite its aim to be a general-purpose grammar, we found
\edibble{}'s semantics ill-suited for HCI experiments. Consequently, \edibble{}
could not express two designs that require decoupling unit specification from
plan construction~\cite{ChanSeatedWIP2024, PottsSweatingDetails2024}. For
example, one study used a degree-nine Latin square with 18
participants~\cite{ChanSeatedWIP2024}. The \edibble{} program generated 18
unique orders rather than the intended Latin square, effectively producing a
block-randomized design. 
Additionally, we found that \edibble{} does not distinguish between
randomization and counterbalancing. 

\touchstone{} has a higher-level of abstraction and provides a set of predefined functions (e.g., \texttt{Latin}, \texttt{Counterbalance}) for canonical HCI designs,
lacking granular or composable primitives. 
We found \touchstone{}'s implementation unfinished. For instance, the
\touchstone2{} paper~\cite{eiselmayer2019touchstone2} claims that \touchstone{}
provides a \texttt{cross} operator, but it is not implemented in the codebase.
Also, \texttt{Random} currently assigns the same order to all participants. We
communicated these findings to the \touchstone2{} authors and evaluated
\touchstone{} assuming \texttt{Random} properly randomized conditions. In its
current state, \touchstone{} could fully express 6 of 15 papers. Among the
papers \dsl{} could express but \touchstone{} could not, three relied on the
\texttt{cross} operator~\cite{WuARTiST2024, DesaiChin2023}, and three were due
to \touchstone{}’s limited support for within-subjects
designs~\cite{shen2024mousering, YuSkinergy2023, Danry2023Explainable}.
\autoref{tab:perPropertyResults} summarizes how each tool supports
experimental-design properties.

\newcommand{\rqExpertConceptualization}{\textbf{RQ1: Existing Practices}}
\newcommand{\rqConceptualMapping}{\textbf{RQ2: Conceptual Mapping}}

\newcommand{\chat}{Wacharamanotham}
\newcommand{\emi}{Tanaka}
\section{Critical Reflections with Experts}~\label{sec:expert-eval} 

Additionally, we conducted critical reflection~\cite{CriticalReflection2019} sessions with the
author of \edibble{} and one author of \touchstone2{}. The following research
questions guided our conversations: 

\begin{itemize}[nosep,leftmargin=*]
    \item \rqExpertConceptualization. What do experts find most important in
    designing and communicating experimental procedures? 

    \item \rqConceptualMapping. In what ways does \dsl{} align with or diverge
    from how experts currently specify, conceptualize, and communicate
    experimental designs?
\end{itemize}

The first author
engaged each expert in a semi-structured interview about important experimental
design practices and how software supports these practices. The
first author presented an example \dsl{} program compared to an equivalent
program in the respective tool (\edibble{} or \touchstone{}). 
Sessions were conducted and recorded over Zoom. Each session lasted between one
and two hours. All study materials are included as supplemental material. Given
the small number of individuals involved in developing these tools, anonymity
was not possible. The experts agreed to going on the record, so we 
identify and attribute quotes to each expert by name.

\subsection{Discussion with \edibble{} Developer} \keyidea{Emi Tanaka is a
statistics professor whose experience designing complex agricultural studies led
her to develop \edibble{}.} She designed \edibble{} to make ``assignment
procedures as explicit as possible'' using a ``model-based'' approach rather
than selecting predefined ``recipes'' (i.e., off-the-shelf experimental
designs). 
In this way, \edibble{} and \dsl{} share
the goal of making structural relationships between experimental designs
explicit. 

\keyidea{Despite this common goal, \emi{} observed that agricultural experiments
use different jargon from human-subjects experiments.}
Specifically, \emi{} noted that, in within-subjects experiments, the unit is the
participant at a given point in time. In \edibble{}, a researcher must
explicitly define the time or position in an order of conditions as a unit. In
contrast, \dsl{} transforms designs with within-subjects variables into an
underlying matrix representation, which captures the implicit unit-time
structure while preserving terminology (syntax) familiar to human-subjects
researchers. Moreover, an additional benefit of \dsl{}'s implementation as
constraint satisfaction over matrices is that it enables extensible analysis
(\autoref{sec:analysis}). 

While the goal of making design choices explicit informed the
abstractions in both \edibble{} and \dsl{}, both had domain-specific
blindspots. Furthermore, \emi{} believed that leaving some details, such as the
definition of a unit, implicit is acceptable when it reflects a community-held
convention. Otherwise, if a DSL aims to capture every low-level detail, a
specification can become too pedantic and ``cause friction'' for researchers to
use. 
 This highlights a social
consideration about how to design DSLs to promote clear communication.

\subsection{Discussion with \touchstone2{} Developer}

Chat \chat{} is an HCI professor who worked on
\touchstone2{} to promote transparent research practices in HCI.

\keyidea{Whereas \dsl{}
derives different experimental designs by layering composable operations,
\touchstone2{} offers a ``recipe-style'' library of pre-defined
techniques (e.g., fully counterbalanced, Latin square).} 
This
composability has downstream consequences: \dsl{} solves constraints over
matrices, enabling it to determine testable effects and compare designs
(\autoref{sec:analysis}). 
\touchstone2{} constructs designs via procedural
scripts and cannot reason about a design's analytical properties. 
Instead, it
focuses on power analysis, which requires additional information about
anticipated effect and sample sizes. \chat{} described a good experimental design as one
that maximizes information given resource constraints, suggesting that
testability, assumptions, and statistical power should all be reasoned about jointly. 
However, existing tools address each separately. 
We discuss this in more detail in \autoref{sec:future-work}.

Initially, \chat{} advocated for higher-level abstractions to ease
specification, but after walking through the \texttt{nest} operation, he changed
his mind, remarking that researchers ``say that we need a better language'' to
describe experiments. He recognized \dsl{} could be a ``candidate language for
discussing experimental design\ldots{} so hiding it from the user may not be the
best idea.'' Although \dsl{} requires more specification than \touchstone2{},
its explicit notation may facilitate more accurate mental models about how
design choices influence experimental outcomes. 

\newcommand{\rqDesignAccuracy}{\textbf{RQ1: Design Success.}}
\newcommand{\rqConceptualClarity}{\textbf{RQ2: Conceptual Clarity.}}
\newcommand{\rqComparativeReasoning}{\textbf{RQ3: Comparative Reasoning.}}

\section{User Evaluation}\label{sec:user-study} 

We conducted six\footnote{Prior work suggests that even five participants can
uncover valuable usability insights~\cite{Jakob1994Usability}.}
think-aloud studies 
to investigate the following research questions:

\begin{itemize} [nosep,leftmargin=*]
    \item \rqDesignAccuracy{}  
    Can researchers use \dsl{} to author and compare experimental designs? What
    challenges or issues do researchers encounter when using \dsl{}? 

    \item \rqConceptualClarity{} What conceptual insights do researchers gain by
    using \dsl{}? Does \dsl{} help researchers think critically about hidden
    assumptions?

    \item \rqComparativeReasoning{} How does \dsl{} influence how researchers
    identify and reason about differences between experimental designs?
\end{itemize}

Through internal message boards and professional contacts, 
we recruited researchers who self-reported prior experience conducting or
designing experiments with human subjects. 
To gain insight into how \dsl{} may impact a wide range
of researchers, we intentionally recruited researchers with different levels of
experimental-design experience. 
All but one participant (P4) were PhD students.
Participants came from a range of research areas, including HCI, psychology,
and clinical trials. Two had extensive self-reported knowledge of terminology
used to describe human-subjects experiments. The remaining participants four had conducted or
designed at least one experiment with human subjects but were unfamiliar with
some terminology (e.g., counterbalance, within-subjects). 


\textit{Procedure.}
After providing informed consent, participants completed a think-aloud lab study
consisting of three phases:

\begin{itemize}[nosep,leftmargin=*]
    \item \textbf{Phase 1: Tutorial.} Participants followed a tutorial covering
    \dsl{}'s key language constructs (e.g., variables, designs, \\
    \texttt{counterbalance}) and terminology (e.g., interaction
    effects). They could access \dsl{}'s documentation throughout.

    \item \textbf{Phase 2: Design specification.} Participants used \dsl{} to
    specify a univariate Latin square from a prose description of an experiment
    in our expressivity evaluation (\autoref{sec:evaluation})
     \cite{Lin2024rambler}. 
     We emphasized
    that there was no single correct answer, as prose descriptions are often
    ambiguous.

    \item \textbf{Phase 3: Design exploration.} Participants explored
    alternative designs for the same research question. They could
    add variables or alter the assignment strategy and were asked to compare at
    least two designs.
\end{itemize}

The study concluded with a semi-structured interview about participants'
experiences designing, analyzing, and comparing designs in \dsl{}, followed by a
closing survey. Researchers were encouraged to think aloud throughout. Sessions
lasted approximately one hour and were conducted and recorded over Zoom.
Participants were compensated \$25 for their time. 
All study materials are included as supplemental material.

\textit{Analysis.}
We analyzed data from system interaction logs, surveys, and interview transcripts. 
In addition, we examined the design specifications and
analyses from each participant, focusing on how \dsl{}'s analysis
informed users' design choices. We conducted a thematic analysis to extract key themes
related to their reasoning strategies from the transcripts. 
The first author coded the
transcripts and resolved discrepancies through discussion with co-authors.

\subsection{Findings}

\subsubsection{RQ1: Researchers can use \dsl{} to specify and iterate on designs
successfully.}  All participants successfully used \dsl{} to implement the
presented design, constructing and comparing between two and six designs. 
Although all participants reported a learning curve with the \texttt{nest}
construct during the tutorial, all came to understand its behavior during the
study through exploration. 
Four participants (P1, P2, P3, P4) initially specified a univariate Latin
square, and two (P5, P6) specified a design with task and interface, suggesting
ambiguity in the design described in the research paper. Three of four
participants who specified a univariate Latin square (P1, P2, P3) questioned
whether task (e.g., long- vs. short-form writing task) should be included as
a variable in the design specification and included task as a variable when
comparing alternatives.


\subsubsection{RQ2: \dsl{} has the potential to make the process of designing
experiments more intentional.} 

\keyidea{Without \dsl{}, participants explained that they
converge on experimental design choices without exploring alternatives due to
short deadline cycles or knowledge barriers.} For instance, P5 described 
previously using a design that ``we knew could limit the takeaways we could
have'' because of a tight deadline cycle. She continued, ``if I
had seen the analysis implications before and explicitly laid out this way,
maybe it would force me to confront the effects of our resource constraints of
time.'' 

\keyidea{\dsl{} lowered the barriers to brainstorming multiple design
alternatives before settling on one.} Participants reported that they often have
a clear idea about a design to implement, informed by previous experiments or
existing protocols. While constructing the design ``isn't hard to write using a
Python script'' (P5) or Excel (P1), participants reported not always being
convinced that the design they had in mind was the most appropriate design for
their context. Thus, constructing a design itself is not necessarily challenging
for the researchers in our study, but thinking through and comparing their
implications is under-supported in existing tools. For instance, P1 shared an
anecdote of their advisor asking them to conduct a study using a within-subjects
design. P1 explained how if their advisor wanted to consider a comparable
between-subjects design, they could simply ``click a button'' in \dsl{} rather
than reconstruct condition orders or implement an entirely new script. In this
way, P1 remarked that \dsl{} is useful for asking ``what-if'' questions that are
otherwise difficult to consider.

P5 and P6 self-reported having the most experience conducting human-subjects
experiments. 
They used \dsl{} to confirm their understanding. 
For instance, when changing interface from a within-subjects to a
between-subjects variable, P6 correctly predicted how \dsl{}'s output would say
that the design with interface as a between-subjects variable should require
more participants. 
P6's thought-process demonstrates that \dsl{}'s output aligned with conceptual
models of experienced researchers while also allowing them to quickly iterate on
and check their understanding.

\subsubsection{RQ3: \dsl{} provides researchers with a language to describe and
make connections across designs.} \keyidea{All participants reported prior
exposure to designs expressible in \dsl{}, but they lacked precise language to
describe them.} For instance, P5, who had five years of experience conducting
HCI experiments and familiarity with all terminology in the pre-screening survey, noted that, using \dsl{}, they reason ``much
more explicitly and [it] forces you to formalize and put words to things.'' While
using the \texttt{nest} operator, she realized she had used a similar procedure
in a prior study. When asked how she had described it at the time, she recalled
that she came up with ``all of the different combinations [of conditions]... and
then multiplying until I matched the number of participants.'' Similarly, P4,
who had experience with clinical trials, reported using diagrams in PowerPoint to
communicate complex designs. Yet, PowerPoint ``is almost too flexible.'' She
reported that they often simplify their design for the sake of communicating the
procedure. These challenges with verbose, procedural descriptions illustrates
how researchers often lack precise, standardized language for common design
choices. Researchers instead must rely on informal prose or figures that, when included in
papers, can be ambiguous. 
Participants found that \dsl{} provides a helpful alternative.

\subsubsection{System Improvements and Other Takeaways}
We
fixed bugs and iterated on \dsl{}'s GUI based on feedback from participants. The
largest change we made was modifying the warning descriptions in \dsl{}'s
analysis panel to use less jargon. 
Half the participants (P3, P5, P6) suggested that \dsl{} is best for researchers
who are already familiar with terminology used in \dsl{} but have limited
experience applying the concepts in designing experiments. They further remarked that
\dsl{} could serve as a teaching tool to help understand design trade-offs. 




\section{Discussion}
Our goal is to improve how researchers communicate and reason about experimental
design. Towards this aim, we formalize experimental assignment procedures in a
grammar, instantiate this grammar in the \dsl{} DSL, and explicate assumptions
underlying designs through a static analysis of \dsl{} programs. Through
expressivity, expert, and user evaluations, we gather evidence that \dsl{}
facilitates communication, reveals hidden structure, and encourages precise
reasoning about experimental design.


\textit{\dsl{} facilitates clear communication of scientific procedures and
assumptions.} \keyidea{\dsl{} promotes transparent research practices by
providing a language of experimental assignment procedures.} Often, experimental
procedures described in prose are verbose and ambiguous
(\autoref{sec:user-study}). In contrast, \dsl{} offers a minimal set of
primitives that underlie both canonical and bespoke designs so that researchers
can specify assignment procedures precisely. As \chat{} remarked in a critical
reflection, our grammar provides a ``candidate language for experimental
design'' (\autoref{sec:expert-eval}). Without a clear language, ambiguous
descriptions make it difficult to challenge assumptions. By mitigating
ambiguity, \dsl{} can promote scientific rigor and facilitate standardized
reporting for replicating experimental findings. 


\textit{Defining \dsl{}'s grammar revealed surprising details and subtle, often
implicit, assumptions about experimental designs.}  We found that experiments
with human subjects adapt designs from experimental design theory to account for
time-based influences. Formalizing counterbalancing in \dsl{} clarified that
Latin squares are useful because they systematically distribute treatments
across blocking variables, whether spatial (as in agriculture) or temporal (as
with human subjects). 
Recognizing implicit assumptions about experimental units and blocking factors
 clarifies that Latin squares are a counterbalancing technique used to account
 for time-based effects in experiments with human subjects. 
In human-subjects experiments the experimental unit can be defined as
 participant-time, with time as the only blocking factor. This framing aligns
 with Fisher’s idea of blocking on orthogonal factors (e.g., row and column). In
 our critical reflection with \emi{}, she remarked that clearly communicating
 details about the participant-time relationship is important for transparent
 research and understanding design decisions (\autoref{sec:expert-eval}). \dsl{}
 captures this implicit unit structure without requiring users to specify
 procedural assignment details. Understanding which designs share similar
 structures can encourage researchers to systematically explore the design space
 while holding essential requirements (e.g., counterbalancing) constant.

\textit{\dsl{} is the first tool to scaffold conceptual trade-offs that
influence which causal queries an experiment can answer.} \dsl{} enables
researchers to reason about design decisions before data collection. As a
result, researchers may no longer need to select suboptimal designs under
deadline pressure, as described by P5 in \autoref{sec:user-study} Instead, we
hope researchers using \dsl{} can reason about what hypotheses their experiment
can test under different assumptions before spending time, money, and energy
conducting experiments. 
\section{Limitations}\label{sec:limitations}

There are two key limitations of the current work.

\textit{Grammar scope.}
\dsl{}'s grammar addresses considerations and uses terminology specific to
human-subjects experiments, especially those found in HCI proceedings. Other
domains have different concerns. For example, as \emi{} explained during a critical reflection session, agricultural trials balance across
spatial plots, and clinical trials involve adaptive stopping rules. A concrete
first step toward generalization is extending \texttt{counterbalance} to balance
conditions across unit attributes (e.g., spatial location) rather than
exclusively across time.

\textit{Unique orders.}
In \autoref{sec:evaluation}, we identified one experiment that no tool could
express. The design assigned multiple orders to each participant and ensured
that no order repeated across participants. 
We could extend \dsl{} to capture this structure by recursively constructing
viable plans, treating orders (i.e., plans) as assignable conditions. As long as
properties of assigned conditions could be derived from properties of assigned
orders, the analysis would remain unchanged. Yet, it remains unclear whether
this added feature would generalize beyond this specific experiment. 

\section{Future Work} \label{sec:future-work}
Developing and evaluating \dsl{} revealed new opportunities for deriving
experimental designs from researchers' hypotheses, 
incorporating statistical power constraints, and 
using prior experiments to inform design choices.

\textit{Deriving designs from hypotheses and assumptions.}
Researchers' hypotheses, domain assumptions, and practical constraints could
more explicitly drive automatic generation of candidate designs. Since many designs can test the
same hypothesis, such a system could 
take as input a high-level hypothesis, set of assumptions, and practical constraints (e.g., budget, time) and 
output a ranked set of options rather than
requiring manual construction. Recent work on formalizing
hypotheses~\cite{suh2022grammarHypo} offers a promising starting point for
mapping causal assumptions to design properties.

\textit{Incorporating statistical power constraints.}
\dsl{}'s analysis focuses on design properties that influence the ability to
test for a causal effect. The number of observations (i.e., participants)
required to detect an effect is also an important consideration. Therefore,
future work should combine the analysis we introduce here with statistical power
analyses, such as those in \touchstone2{}~\cite{mackay2007touchstone}, to further support researchers navigate practical considerations in experiments. 
For example, if two designs
surface identical assumptions but differ in the number of participants required,
\dsl{} could recommend the more efficient option within a researcher's budget 
(\autoref{sec:expert-eval}).

\textit{Supporting iterative study design.}
\dsl{} treats each experiment as a standalone specification. In practice,
researchers iteratively refine procedures across a sequence of studies. Future
work could take prior design specifications and analytical results as input to
suggest modifications. For example, when a prior study reveals significant order
effects, the system might suggest a between-subjects design.

\section{Conclusion}
\dsl{} implements a grammar of experimental assignment, translating core
concepts into logical expressions over matrices. 
\dsl{}'s composable grammar enables analysis of testable
causal effects. Our work supports a broader vision of
improving the validity and understanding of experiments as well as fostering more
precise communication among researchers. \dsl{} is publicly available but \textit{not
anonymous} on \texttt{pip}: \texttt{pip install planet-dsl}. Additional information and \textit{anonymous} source code are
available at \url{https://anonymous.4open.science/r/planet-lang-interface-5D70}.

\bibliographystyle{ACM-Reference-Format}
\bibliography{adamc, emjun, london}

\appendix
\clearpage{}
\section{Appendix}~\label{sec:appendix}

\begin{definition}[Kronecker product~\cite{Dufour2018}]
    Let A be an $ n \times m$ matrix and B a $k \times p$ matrix. The Kronecker
    product between A and B is the $kn \times mp$ block matrix \\
    $A \otimes B = 
    \begin{bmatrix}
        A_{11}B & \cdots & A_{1m}B\\
        \vdots & \ddots & \vdots\\
        A_{n1}B & \cdots & A_{nm}B\\
    \end{bmatrix}
    $
  \end{definition}

\NestComparisonFigure{}
\perPaperResults{}



\begin{table*}[t]
    \centering
    \setlength{\tabcolsep}{2pt}
    \caption{\textbf{Papers with relevant experimental-design keywords}. The
    percentage is relative to the total number of papers with relevant
    keywords. The final column indicates the number of papers
    with relevant keywords in the conference
    proceedings. The labels are not mutually exclusive.}
    ~\label{tab:sample}
    \begin{tabular}{lrrrrr|r}
        \toprule
         & user study & within-subjects & between-subjects & Latin
         square & counterbalance & total papers
         \\
        \midrule
        CHI & 161 (21.32\%) & 471 (62.38\%) & 254 (33.64\%) & 107
        (14.17\%) & 294 (38.94\%) & 755  \\
        UIST & 11 (10.89\%) & 78 (77.23\%) & 9 (8.91\%) & 20 (19.80\%) & 64 (63.37\%) & 101  \\
        \bottomrule
    \end{tabular}  
\end{table*}

\begin{table*}[htbp]
    \centering
    \caption{\textbf{Execution time of each \dsl{} program from \autoref{sec:evaluation}
    in seconds}. All programs were run on a MacBook Pro (M4 Pro, 24 GB). All programs completed in under 33 seconds, with 14 of the 15
    finishing in under 1 second.
    }
    \label{tab:execution-times}
    \begin{tabular}{p{0.75\textwidth}c}
    \toprule
    \textbf{Paper} & \textbf{Execution Time (s)} \\
    \midrule
    ARTiST: Automated Text Simplification for Task Guidance in Augmented Reality (Exp 1)~\cite{WuARTiST2024} & 0.214 \\
    ARTiST: Automated Text Simplification for Task Guidance in Augmented Reality (Exp 2)~\cite{WuARTiST2024} & 0.520 \\
    Don't Just Tell Me, Ask Me: AI Systems that Intelligently Frame Explanations as Questions Improve Human Logical Discernment Accuracy over Causal AI explanations~\cite{Danry2023Explainable} & 0.205 \\
    MouseRing: Always-available Touchpad Interaction with IMU Rings~\cite{shen2024mousering} (Exp 1) & 32.297 \\
    MouseRing: Always-available Touchpad Interaction with IMU Rings~\cite{shen2024mousering} (Exp 2) & 1.660 \\
    OK Google, Let's Learn: Using Voice User Interfaces for Informal Self-Regulated Learning of Health Topics among Younger and Older Adults~\cite{DesaiChin2023} & 0.208 \\
    Understanding Perception of Human Augmentation: A Mixed-Method Study~\cite{Villa2023Augmentation} & 0.205 \\
    Rambler: Supporting Writing With Speech via LLM-Assisted Gist Manipulation~\cite{Lin2024rambler} & 0.201 \\
    Seated-WIP: Enabling Walking-in-Place Locomotion for Stationary Chairs in Confined Spaces~\cite{ChanSeatedWIP2024} & 0.341 \\
    Shaping Compliance: Inducing Haptic Illusion of Compliance in Different Shapes with Electrotactile Grains~\cite{jingu2024shaping} (Exp 2) & 1.104 \\
    Shaping Compliance: Inducing Haptic Illusion of Compliance in Different Shapes with Electrotactile Grains~\cite{jingu2024shaping} (Exp 1) & 0.207 \\
    Silver-Tongued and Sundry: Exploring Intersectional Pronouns with ChatGPT~\cite{Fuji2024GPT} & 0.206 \\
    Skinergy: Machine-Embroidered Silicone-Textile Composites as On-Skin Self-Powered Input Sensors~\cite{YuSkinergy2023} & 0.210 \\
    Sweating the Details: Emotion Recognition and the Influence of Physical Exertion in Virtual Reality Exergaming~\cite{PottsSweatingDetails2024} & 0.279 \\
    \bottomrule
    \end{tabular}
\end{table*}

\clearpage

\subsection{Case Studies with Researchers}\label{sec:case-studies} 

We conducted three exploratory case studies to better understand how
different researchers interact with \dsl{} and reason about experimental design. 
The following research questions guided the evaluation: 

\newcommand{\rqDesignAccuracyCS}{\textbf{RQ1: Design Success.}}
\newcommand{\rqConceptualClarityCS}{\textbf{RQ2: Conceptual Clarity.}}
\newcommand{\rqFutureWorkflowCS}{\textbf{RQ3: Future Workflows.}}

\begin{itemize}
    \item \rqDesignAccuracyCS{}  
    Can researchers use \dsl{} to
    author experimental designs successfully? What challenges or issues do researchers
    encounter when using \dsl{}? What desired or missing features come up?

    \item \rqConceptualClarityCS{} What conceptual insights do researchers gain by
    using \dsl{}? Does \dsl{} help reveal or clarify experimental design
    concepts?

    \item \rqFutureWorkflowCS{} How, if at all, do researchers envision
    integrating \dsl{} into their existing 
    workflows? 

\end{itemize}

We recruited researchers with self-reported experimental-design experience and
Python proficiency through internal message boards and professional contacts. To
gain insight into how \dsl{} may impact a wide range of researchers, we intentionally recruited
researchers with diverse experimental-design training.


\paragraph{Procedure.} 
After consenting to participate in the study, researchers followed a tutorial
and read documentation about \dsl{}. The tutorial was a computational notebook
that walked researchers through how to use key language constructs in \dsl{}.
The documentation provided more details about each construct and its
implementation. Next, researchers installed \dsl{} in the notebook environment via \texttt{pip}.
Researchers then specified their designs in a notebook that was prepopulated with an empty
template \dsl{} program. They continued working with \dsl{} until they
expressed being content with their programs. Afterwards, the first author
engaged each researcher in a semi-structured interview about the design process and how it compared to prior experiences. Finally,
the participants completed a closing survey about their experience with \dsl{}.

Researchers were encouraged to think aloud throughout the study. 
The first author answered any of their questions but encouraged them to work independently. 
Sessions were conducted and recorded over Zoom.
Each session lasted between one and two hours.
All study materials are included as supplemental material.

\begin{figure*}[t]
    \centering
    \begin{subfigure}[c]{0.47\linewidth}
      \centering
      \includegraphics[width=\linewidth]{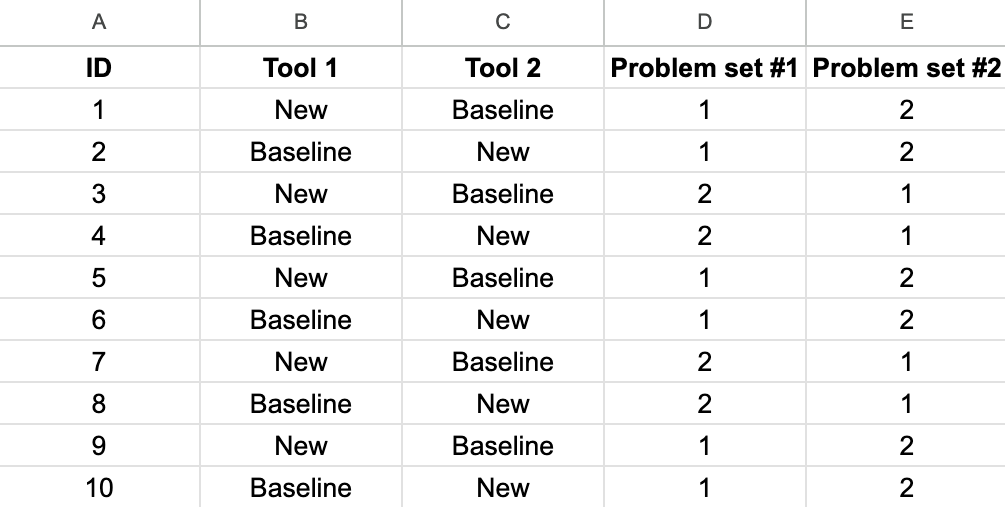}
      \caption{}
      \label{fig:a}
    \end{subfigure}
    \hspace{0.05\linewidth}
    \begin{subfigure}[c]{0.4\linewidth}
      \centering
      \resizebox{!}{0.19\textheight}{%
      \lstset{
        basicstyle=\ttfamily\scriptsize,
        frame=single,
        language=Python,
        numbers=none,
        xleftmargin=0pt,
        framexleftmargin=0pt,
    }
\begin{lstlisting}[language=Python, style=planetstyle]
tools = ExperimentVariable(
    name = "tools",
    options=["New", "Baseline"]
)

tasks = ExperimentVariable(
    name = "tasks",
    options=["1", "2"]
)

participants = Units(14)

design = (
    Design()
      .within_subjects(tools)
      .counterbalance(tools)
      .within_subjects(tasks)
      .counterbalance(tasks)
      .num_trials(2)
)

assignment = assign(participants, design)
print(assignment)
\end{lstlisting}
        }
      \caption{}
      \label{fig:b}
    \end{subfigure}
    \caption{\textbf{R1’s intended experimental assignment and \dsl{} program.} (a) The original
    spreadsheet R1 had previously used to manually construct experimental
    assignments for their study. (b) The
    \dsl{} program for R1's experiment, which produces the same set of orders
    and correctly represented R1's intended
    design (\autoref{subsec:R1}). To preserve anonymity, we renamed the tools in
    the program and spreadsheet.}
    \label{fig:SpreadsheetComparison}
  \end{figure*}

\subsubsection{Case Study 1: Replicating a Complex Study Design}~\label{subsec:R1}
R1, a software-engineering PhD student, had previously conducted and published a complex
within-subjects user study to evaluate a new system. They wanted to replicate that study's design in \dsl{}. 

\paragraph{Previous experiment.}
R1 conducted an experiment to evaluate how a new system influences users' ability
to identify dangerous data flow, a process known as taint analysis. 
The study involved 14 participants.
Each participant was required to complete two trials. 
Each trial involved using one taint-analysis tool to answer a set of questions. 
Each participant was required to use both tools and answer two sets of questions.

R1, who had previously designed two experiments reported in published papers, was accustomed to designing them in spreadsheets. 
During the study, R1 opened the spreadsheet for this experiment (\autoref{fig:SpreadsheetComparison}). 
The spreadsheet contained four columns: two for ``task order'' and two for
``problem set order.''
R1 explained that counterbalancing each variable was important for
answering their research question. However, because they recruited only 14 participants,
it was not feasible to assign fully counterbalanced orders with all four
conditions. 




\paragraph{Using \dsl{}.} To replicate the above study's design in \dsl{}, R1
started by creating a \texttt{multifact} variable to combine tool and task
conditions. Then, they counterbalanced the combined values, using
\texttt{counterbalanced}. The result, as expected, was to assign four
conditions to each participant. However, R1 only wanted to assign two. To
correct the behavior, they reduced the number of trials per participant by adding
\texttt{num\_trials(2)} to the \texttt{Design}. Although closer to the desired
output, this program could assign participants the same tool or task in both trials,
which is something R1 wanted to avoid. 
%
After the interviewer reminded R1 that properties can be applied to individual variables, R1 defined
task and tool as separate within-subjects variables (\texttt{within\_subjects}) and applied counterbalancing (\texttt{counterbalance})
to each. This final adjustment resulted in a design that matched their prior study
(\rqDesignAccuracyCS). Unlike R1's procedural spreadsheet, the \dsl{} program
concisely specified that both task and tool were counterbalanced, with two
conditions per participant (\autoref{fig:SpreadsheetComparison}). R1's \dsl{} program was declarative and made R1's design goals
explicit. 

When asked about how \dsl{} compares with their previous experiences designing
experiments, R1 stated that \dsl{} offered ``the freedom to explore alternatives
without consuming too much time.'' For example, R1 explained that, when manually
designing their experiment, they attempted a fully counterbalanced design with
all four combined conditions. This resulted in more plans (24) than the number
of participants (14). To resolve the mismatch, they counterbalanced each variable
independently and combined the orders, resulting in the four orders they
used in the experiment. In the interview, R1 explained that
\texttt{limit\_plans} changed the way they thought about counterbalancing
conditions. Specifically, they realized they could have counterbalanced all four
combined conditions and limited the number of plans to four. The result is a multifactorial Latin
square, a design R1 had little exposure to. They were not aware of this
possibility before using \dsl{}, which made them reflect on alternative design
choices for future studies. This reaction
indicates that \dsl{}'s composable grammar encouraged R1 to explore designs with
their desired properties without formal training in experimental design
(\rqConceptualClarityCS). Furthermore, R1’s experience shows how \dsl{} enables
experienced researchers to explore alternative designs rapidly while ensuring
critical properties like counterbalancing.

R1 believed \dsl{}'s ``fast-feedback'' could ``save [them] a lot of time in
meetings'' by allowing them to explore experimental-design alternatives on their
own before engaging with collaborators. R1 hoped to use this extra time with
collaborators to ``actually talk about the [primary] research tasks themselves.'' R1
plans on using \dsl{} to design and conduct their next user study
(\rqFutureWorkflowCS), which they anticipate to be within the next year.

\subsubsection{Case Study 2: Applying Experimental-Design Theory}
R2, an undergraduate statistics major, described
their knowledge of experimental design as ``mostly conceptual'' from their
coursework. 
R2 was interested in using \dsl{} to brainstorm experimental designs for future research projects. 

\paragraph{Using \dsl{}.} R2 planned an experiment assessing the effect
of sleep and nutrition on athletic performance. First, R2
defined sleep and nutririon as \texttt{ExperimentVariable}s and specified 20 participants. Referencing the tutorial, R2
used \texttt{multifact} to construct a within-subjects, multifactorial,
counterbalanced design. Those constraints would result in a fully counterbalanced design,
requiring many more partipants than the amount specified. The unintended outcome introduced R2 to
practical considerations when designing experiments. 

Looking at the example of a Latin square in the tutorial, R1 asked a clarifying
question about the function of \texttt{limit\_plans}. Attempting to preserve
counterbalancing while accounting for a limited pool of participants, R2 limited
the number of plans to 20. The \dsl{} program errored because there must be a
multiple of six plans to enforce counterbalancing. Iterating on the design, R2
limited the number of plans to 24, resulting in a set of viable plans. \dsl{}
output a trial table indicating that they should recruit four more participants
to satisfy counterbalancing, a much smaller gap than the fully counterbalanced
design. 

Interestingly, R2's design was neither a Latin square nor a full-counterbalanced
design. Yet, it satisfied R2's desire to counterbalance conditions. Importantly,
this design is difficult to communicate because it is not a ``canonical'' design
we have seen in statistics courses. \dsl{} helped bridge this gap for R2.

R2 observed ``their design [the program] was pretty clear... it's not too
complex because I just have two variables'' (\rqDesignAccuracyCS). The
experimental designs they explored in coursework were ``on a smaller scale,''
designed using ``brute force.'' R2 further explained they ``would lay out the permutations for all of
them.'' Like R1, R2 found that \dsl{} provides a declarative alternative
to time-consuming, manual approaches. Additionally, R2 thought \dsl{} was most
useful for complex designs. Specifically, they identified
\texttt{counterbalance} and \texttt{multifact} as helping them move
beyond ``mostly basic designs.'' Overall, R2's
iterative development illustrates how \dsl{} can support novice researchers in
applying canonical experimental design concepts in more complex settings while
accommodating practical constraints (\rqConceptualClarityCS).

\subsubsection{Case Study 3: Structured Exploration of Adaptive Designs}
R2, an HCI PhD student working on health-sensing tools, had previously conducted
two experiments. 
Presently, R3 was planning a brand new adaptive study to assess the impact of a prompt
intervention on stress. 
Adaptive experiments are out-of-scope for \dsl{}, but R3 remained interested in
expressing variations of their design.


\paragraph{New experiment.}
R2 shared a Google Doc describing their experiment. It spanned
nearly an entire page. Over
several days, a physiological sensor would detect stress responses in
participants, triggering either a generic or tailored intervention. Because user
behavior determined if and when a condition was assigned, participants could
receive differing numbers of conditions at different time
intervals.\footnote{Technically, this example is considered a \textit{microrandomized
trial (MRT)}, an adaptive design where participants are randomized at many
points of an experiment~\cite{Klasnja2015Microrandomization}.}
R2's document exemplified how experimental designs often require verbose, dense
text to capture the details. 

\paragraph{Using \dsl{}.} 
R3 was able to quickly and correctly apply \dsl{}’s operations to define a
simplified version of their experiment, where each participant received only one
condition (\rqDesignAccuracyCS). Within 30 seconds and without any assistance, R3
specified a \texttt{Design} and applied \texttt{between\_subjects} to the
intervention \texttt{ExperimentVariable}. The result was a fully randomized
between-subjects design. 
R3 remarked that \dsl{} ``would have
been really helpful'' compared to previous experiences manually running random
number generators (\rqFutureWorkflowCS). Similar to R1 and R2, R3 found that 
\dsl{}'s declarative language relieved them from monotonous procedural work in coming up with experimental assignments. 

Building off this experimental design, R3 went on to represent repeated assignments across multiple participants. 
R3 modified
their earlier design by replacing \texttt{between\_subjects} with
\texttt{within\_subjects}. They then attempted to increase the number of trials
using \texttt{num\_trials}, but the attempt failed because conditions in within-subject
designs do not repeat by default. After the first author reminded R3 of the \texttt{nest} operator, R3 successfully replicated conditions. 

The resulting \dsl{} program conveyed that their design (i) repeated assignments
over time, (ii) assigned more than one instance of each condition to
participants, and (iii) randomized condition assignment at each trial
(\rqDesignAccuracyCS). \dsl{} helped R3 see parts of their adaptive
experiment in terms of canonical experimental-design language. Previously, in the Google Doc, R3 had described ``within-subjects
analyses,'' but the design's within-subjects structure remained implicit.
Now, in the \dsl{} program, the within-subjects nature of the experiment was explicit. 
Moreover, R3's experience demonstrates that while 
\dsl{} was not originally designed to capture adaptive experiments, \dsl{} can
in fact express core design features. To represent uneven trials and the
relationship between user behavior and assignment decisions, we need to extend \dsl{}'s
grammar. 

After successfully defining a fully randomized, within-subjects design with
repeating conditions, R3 explored the effect of counterbalancing. They reported,
``ideally we want balanced conditions, but we don’t make the extra effort in
[our] field studies.'' By adding a single \texttt{counterbalance} operation,
they could rapidly compare the counterbalanced version to their previous design.
Without \dsl{}, directly comparing the designs requires
manual scrutiny of the output.

\subsubsection{Takeaways}
All three researchers 
expressed benefitting from
\dsl{}'s composable
grammar to specify simple-to-complex designs declaratively. Using \dsl{} would
replace the ``time-consuming'' tasks currently involved in designing
experiments: manual construction using spreadsheets (R1), repeatedly
running random number generators (R2), and permuting all
possible orders (R3). 
Furthermore, R1 reported that they verified
designs ``by getting it reviewed by other people,'' suggesting that
effective communication is crucial to ensure valid assignment procedures. This
process is currently manual and error-prone with existing tools and methods for
designing experiments. Instead, \dsl{} programs could provide an intermediate representation,
facilitating communication among collaborators. 

All researchers initially found \texttt{cross} and \texttt{nest} operations to be
challenging. R3 reported that ``it took a few more minutes to
understand'' and eventually became ``intuitive... even in first time use.''
R1 stated that it was a matter of ``reading the documentation.'' 

We observed that all researchers adopted an iterative approach to designing
their experiments with \dsl{}, removing, adding, and replacing operations to
designs. This allowed researchers to explore alternatives rapidly while
preserving important design properties, such as counterbalancing or limiting the
number of conditions. For instance, both R1 and R2 desired counterbalanced
conditions, but a small participant pool limited their choices. For R1, \dsl{}
helped identify a multifactorial Latin square as a valid design option,
something R1 overlooked during the manual design process. R3 similarly used
\dsl{} to reason about properties of an adaptive experiment.
These examples show how rapid exploration of alternatives can improve conceptual
clarity 
and allow researchers to go beyond textbook designs. In fact, all researchers
reported that \dsl{} improved their understanding of experimental design in the
closing survey.

\end{document}